\documentclass[aps, notitlepage, pre, 12pt, superscriptaddress]{revtex4-2}
\usepackage{graphicx, natbib, titlesec, paralist, amsmath,amssymb, color}
\usepackage{url} 
\usepackage{lineno}
\usepackage{soul}
\usepackage[english]{babel}

\usepackage{xr}
\makeatletter
\newcommand*{\addFileDependency}[1]{
  \typeout{(#1)}
  \@addtofilelist{#1}
  \IfFileExists{#1}{}{\typeout{No file #1.}}
}
\makeatother
\newcommand*{\myexternaldocument}[1]{%
    \externaldocument{#1}%
    \addFileDependency{#1.tex}%
    \addFileDependency{#1.aux}%
}
\myexternaldocument{supporting_information}

\begin{document}

\title{Diversity of growth rates maximizes phytoplankton productivity in an eddying ocean}


\author{Mara A. Freilich}
\email{maraf@mit.edu}
\altaffiliation[Current address: ]{Scripps Institution of Oceanography, La Jolla, CA}
\affiliation{Earth, Atmospheric and Planetary Sciences, MIT, Cambridge, MA 02139}
\affiliation{Woods Hole Oceanographic Institution, Woods Hole, MA 02543}
\author{Glenn Flierl}
\affiliation{Earth, Atmospheric and Planetary Sciences, MIT, Cambridge, MA 02139}
\author{Amala Mahadevan}
\affiliation{Woods Hole Oceanographic Institution, Woods Hole, MA 02543}

\date{September 21, 2021}

\begin{abstract}
In the subtropical gyres, phytoplankton rely on eddies for transporting nutrients from depth to the euphotic zone. But, what controls the rate of nutrient supply for new production? We show that vertical nutrient flux depends both on the vertical motion within the eddying flow and varies nonlinearly with the growth rate of the phytoplankton itself. Flux is maximized when the growth rate matches the inverse of the decorrelation timescale for vertical motion. Using a three-dimensional ocean model and a linear nutrient uptake model, we find that phytoplankton productivity is maximized for a growth rate of 1/3 day$^{-1}$, which corresponds to the timescale of submesoscale dynamics. Variability in the frequency of vertical motion across different physical features of the flow favors phytoplankton production with different growth rates. Such a growth-transport feedback can generate diversity in the phytoplankton community structure at submesoscales and higher net productivity in the presence of community diversity.
\end{abstract}

\maketitle

\section*{Plain Language Summary}
The productivity of phytoplankton is the result of a two-way coupling between the rate of phytoplankton growth and nutrient supply by physical mechanisms. A three-dimensional model of an oceanic eddy flow field maximizes its nutrient flux to the sunlit surface layer when the average phytoplankton growth rate is approximately 1/3 day$^{-1}$, which is consistent with observations. Different features of the flow field experience variable frequencies in the up/down motion and favor different phytoplankton growth rates to maximize new production, providing a way to generate diversity in the community composition of phytoplankton. Since changing the grid resolution of ocean models alters the vertical velocities, the biological growth rates also ought to be altered to represent the appropriate physical-biological coupling.

%
%

%


%
%
%
%

\section{Introduction}
The diverse community of phytoplankton and the heterotrophic ecosystem that it supplies affects the depth and efficacy of ocean primary production, as well as the cycling of carbon and other elements. Most of the world's ocean is oligotrophic (depleted of nutrients) in the sunlit (euphotic) layer where they are taken up by phytoplankton, but nutrient concentration increases with depth below the euphotic layer. In such regions, the production of phytoplankton relies on the physical transport of nutrient-enriched water from depth to the euphotic zone where light enables photosynthesis \cite{mcgillicuddy1998influence}. Over long spatial and temporal scales, the system is balanced such that the rate of export of organic matter is determined by the rate of nutrient input that contributes to photosynthetic carbon fixation \cite{falkowski1998biogeochemical,ducklow2001upper}. 

The upward transport of nutrient-rich water occurs via a range of mechanisms, including surface boundary layer turbulence, wind-driven upwelling, coastal upwelling, eddy uplift and frontal instabilities \cite{denman1983time,lipschultz2002new}. In the pelagic ocean, the physical supply of nutrients induced by vertical advection associated with fronts and eddies \cite{jenkins1985seasonal} is thought to limit the rate of {\em new} production, which is the rate of phytoplankton production fueled by a fresh supply of macronutrients from outside the euphotic layer. Vertical velocities are typically 10$^{-3}$ to 10$^{-4}$ times smaller than the 
horizontal velocities associated with ocean currents and eddies on scales of 1--100~km. 
But, submesoscale dynamics, associated with strong vertical vorticity (of the order of the planetary vorticity $f$) 
 can result in vertical velocities of $\mathcal{O}$(100)~m/day on spatial scales $\mathcal{O}$(1~km). These rapid vertical motions are thought to be particularly influential for phytoplankton growth.

Besides physical processes, the gradient in the mean nutrient distribution and the anomalies in concentration from the mean distribution also affect nutrient transport.  
Physical transport, as well as  biological and chemical processes that create sinks (or sources) for nutrients, affect their distribution.
The spatial heterogeneity of reactive (biogeochemical) tracers in the ocean is  dependent on  the Damk\"{o}hler number, $Da$, which is the ratio of the advection to reaction timescales \cite{lopez2001chaotic,abraham1998generation,mahadevan2000modeling}. It controls the nonlinear relationship between reaction products and physical decorrelation timescales. 
 $Da$ affects the efficiency of transport \cite{mahadevan2002biogeochemical,pasquero2005impact,smith2016effects}, which is optimized for $Da=\mathcal{O}(1)$. 

Hence, the nutrient supply rate cannot be assessed from physics alone; it also depends on the rate of nutrient uptake by phytoplankton, which is closely related to the growth rate, a physiological characteristic that depends on the kind of phytoplankton, its size, the ambient temperature, light and nutrient availability.
Ocean ecosystems are highly heterogeneous with phytoplankton cell sizes and growth rates varying by orders of magnitude \cite{laws2013evaluation}.
This diverse range of phytoplankton has a wide range of growth rates that are both affected by and affect the nutrient supply rate. 
\section{Nutrient supply and uptake}
Oceanic biogeochemical tracers typically have a strong depth dependence, with relatively weak lateral variability due to density stratification and rotation, which inhibit vertical movement of water, and the dependence of phytoplankton growth on light. In the oligotrophic, subtropical oceans, the area- and time-averaged mean vertical profile of nutrient $N_0(z)$ is depleted in the near-surface euphotic layer (upper 100~m), and has a strong vertical gradient (nutricline) in the region of strong density stratification (pycnocline) approximately between 100--500~m (Fig.~S10). The nutrient concentration is altered by advection and nutrient uptake (or resupply), which we model as a linearized resource-limited source/sink function $f(N) = - \lambda ( N- N_0(z) )$ \cite{abraham1998generation,lopez2001chaotic}. This function is a linearization of logistic growth or decay near carrying capacity (SI Text 2.1). Our simple model for nutrient 
\begin{equation}
    \frac{\partial N}{\partial t}+   \nabla \cdot ({\bf u} N) =   - \lambda ( N- N_0(z) ) 
    \label{eq:original}
\end{equation}
accounts for transport by the ocean velocity field ${\bf u} = (u,v,w)$ and the uptake (resupply) of nutrients \cite{mahadevan2000modeling} and averages over several aspects of the ecosystem and the interactions of its components due to grazing, mortality, detritus production, and bacterial remineralization. Here, $\lambda$ is the rate of uptake or resupply (per unit time). For practical purposes, we take $\lambda$ to be depth-independent in our model and encode all the depth dependence in $N_0(z)$, but in a general model $\lambda$ could vary in space and time, or be a function of temperature. 
By defining the nutrient anomaly $N'(x,y,z,t)  \equiv N(x,y,z,t) - N_0(z)$ and taking the horizontal area-average, denoted by $\langle ~ \rangle$ of (\ref{eq:original}), over a region with no large-scale horizontal gradients in $N$, we are left with a balance between vertical transport and nutrient uptake or resupply, because the spatial average of the vertical velocity ($w$) vanishes, $\langle w \rangle = 0$ and $ \partial_x\langle u  N' \rangle =  \partial_y\langle v N' \rangle = 0$, such that
\begin{equation}
  \langle  \partial_t N' \rangle + \langle \partial_z ( w  N' )  \rangle =   -  \langle \lambda  N' \rangle.
    \label{eq:anomaly}
\end{equation}
Nutrient anomalies are generated by the vertical advection of nutrient and restored to the equilibrium profile at a rate $\lambda$.  Where $N' > 0$, nutrient is consumed, as by the new production of phytoplankton.  Where $N' < 0$, nutrient is restored to its equilibrium profile, and the resupply represents remineralization.  Time-averaged over a the long term (several months) we expect a steady state in which the nutrient flux  at the base of the euphotic layer ($z=-z_e$) is balanced by the nutrient consumption above it  $\langle w N' \rangle_{z_e} = - \int_{-z_e}^0 \langle \lambda N' \rangle dz$.  In a time- and area-averaged sense, the net uptake (resupply) is equivalent to the net community production (NCP), which over the long term, is balanced by the net export and is often, as it is here, conceptualized in terms of first order rate kinetics \cite{woodwell1968primary,emerson2014annual}.

\begin{figure}
\centering
\includegraphics[width=0.8\textwidth]{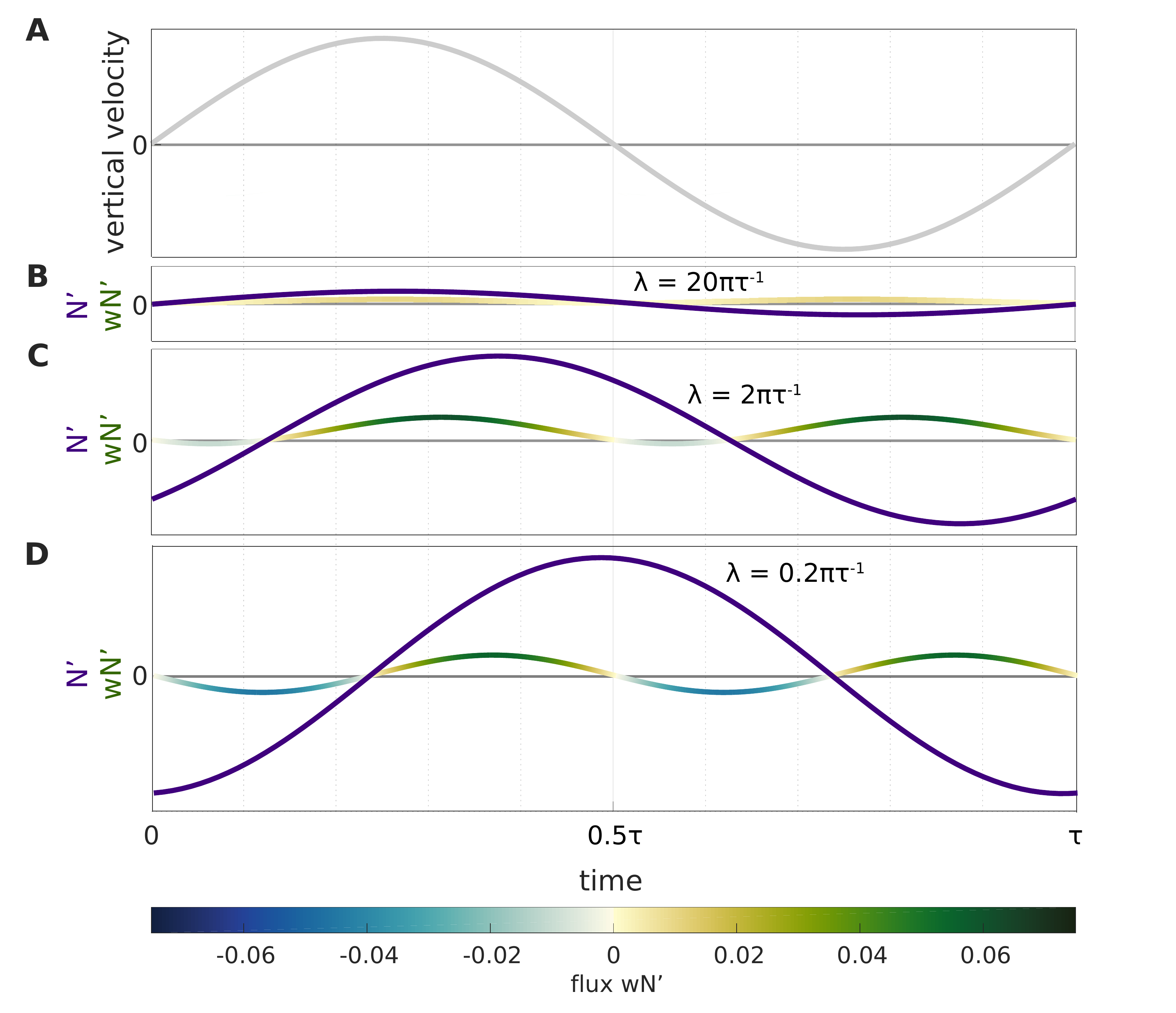}
    \caption{Phase and amplitude relationships between an idealized, oscillatory vertical velocity (panel A), and nutrient anomaly ($N'$ in purple) and nutrient flux ($wN'$ in green (positive) and blue (negative)) over one time period $\tau$. B,C and D show $N'$ and $wN'$ for 3 different values of uptake rate $\lambda$. The time-integrated flux $\overline{wN'}$ is maximized when $\lambda=\frac{2\pi}{\tau}$ (panel C) as described in Section~\ref{sec:sinusoidal}}
    \label{fig:schematic}
\end{figure}

\subsection{Dependence of flux on the frequency of vertical velocity and uptake rate}
\label{sec:sinusoidal}
The dependence of the nutrient flux on the covariance between $w$ and $N'$ can be demonstrated conceptually (Fig.~\ref{fig:schematic}) by using an oscillating, sinusoidal, vertical velocity $w=w_0  \sin (\frac{2\pi}{\tau} t)$ with period $\tau$ and angular frequency $\frac{2\pi}{\tau}$. When the uptake rate $\lambda \gg \tau^{-1}$ (fast uptake), the nutrient anomaly is in phase with the vertical velocity, but its magnitude is small because nutrients are consumed immediately (Fig.~\ref{fig:schematic}B). Consequently, the magnitude of the flux is small. By contrast, when $\lambda \ll \tau^{-1}$ (slow uptake), nutrients remain unconsumed and the nutrient anomaly is large, but out of phase with the vertical velocity (Fig.~\ref{fig:schematic}D). Because the nutrient anomaly and vertical velocity are out of phase, there are times of positive and negative flux, which, when averaged in time  (and denoted by an overbar), results in a small net vertical flux $\overline{w N'}$. The net nutrient flux is maximized between those two extremes, when $\lambda = \frac{2\pi}{\tau}$
and the nutrient anomalies have a larger magnitude relative to the fast growth case and a $45^\circ$ phase shift relative to the vertical velocity (Fig.~\ref{fig:schematic}C). As a consequence, the net positive flux is maximal when averaged over a time period (SI Text 2.3). 

\begin{figure}[htb]
\centering
\includegraphics[width=1\linewidth]{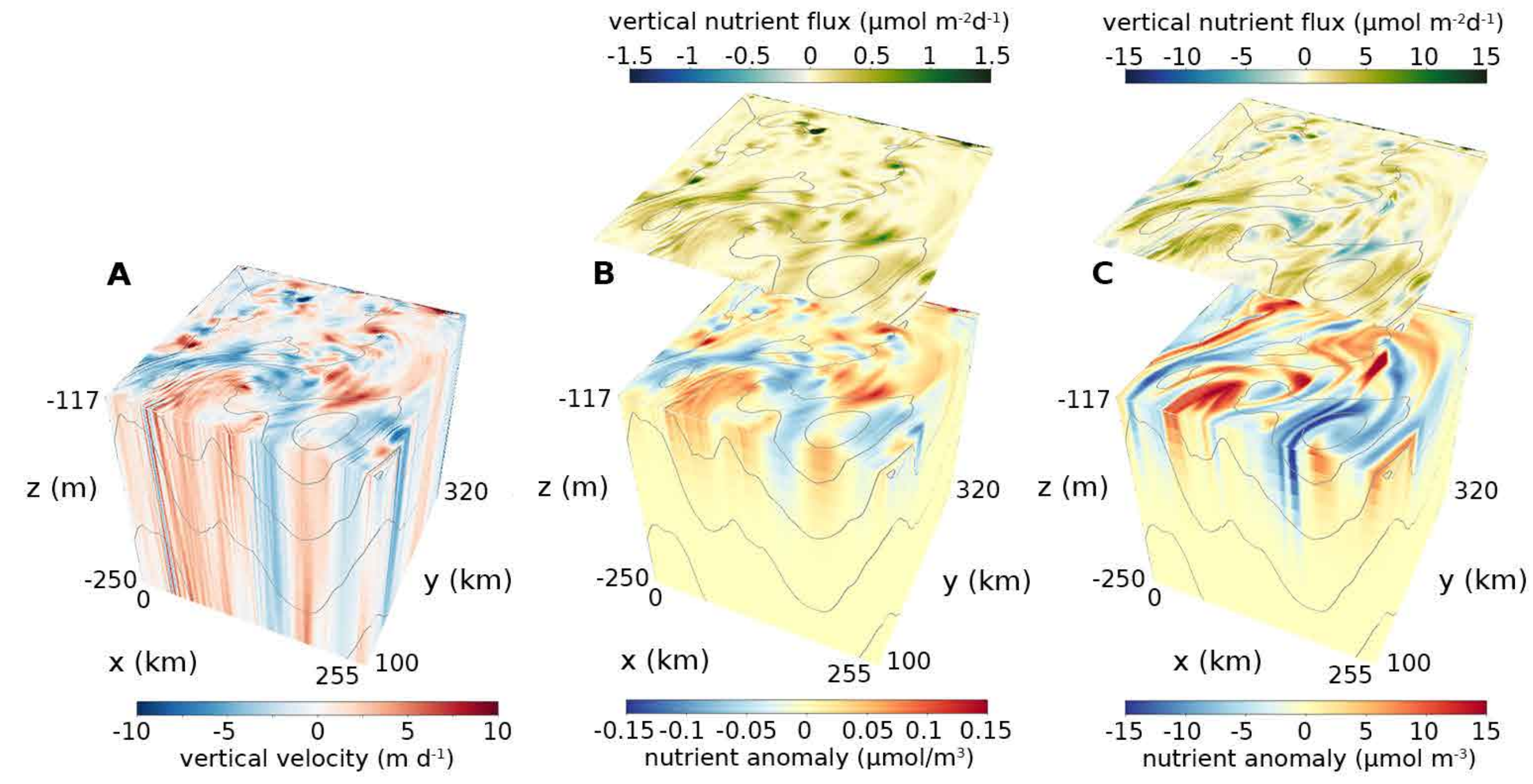}
    \caption{Snapshots of vertical velocity, nutrient anomaly, and vertical nutrient fluxes on model day 135 (A) Vertical velocity at $z = -117$~m. (B and C) Lower cube is nutrient anomaly, the upper panel is the vertical nutrient flux at $z = -117$~m with uptake rates (B) $\lambda = 0.015$~day$^{-1}$ and (C) $\lambda = 1.5$~day$^{-1}$}
    \label{fig:3D}
\end{figure}

\subsection{Nutrient fluxes in an oligotrophic eddy field}
Vertical motions occurring on a wide range of spatial and temporal scales \cite{freilich2019decomposition} contribute to biological tracer fluxes, resulting in more complex patterns of eddy correlation than in a simple oscillatory flow field. We use a numerical ocean model to examine the dependence of the vertical nutrient flux on the uptake rate $\lambda$ in the fully developed, nonlinear oceanic regime.  In such a flow field, we consider the characteristic timescale $\tau$ of the vertical velocity to be the decorrelation timescale of Lagrangian water parcels as they move up and down across the nutricline.
A three-dimensional (3D), high-resolution (1~km horizontal and variable vertical resolution) numerical ocean model is used to simulate a mid-latitude eddy field that is based on hydrography from the western North Atlantic subtropical gyre. The model is a periodic channel 256 $\times$ 320 km in extent and 1~km deep. The developed flow field represents characteristics of the subtropical eddy field as described in \cite{freilich2019decomposition}.  
The strongest vertical velocities are at horizontal spatial scales of a few kilometers and occur on the edges of $\sim$100~km scale mesoscale eddies (Figure \ref{fig:3D}A). We couple this physical model to the simplified nutrient model (\ref{eq:original}) and use a representative regional mean as the equilibrium nutrient profile $N_0(z)$ with the nutricline centered at 115 meters (Fig. S10). 
The maximum new production in the model is centered at 115 meters depth, the location of the maximum gradient in the nutrient profile.  We vary the uptake rate $\lambda$ between 0.005--50 day$^{-1}$, which encompasses the observed range of  phytoplankton growth rates in nutrient-limited mid-latitude subtropical gyres \cite{laws2013evaluation}. 

On average, water parcels ascending through the nutracline into the euphotic zone will carry a positive nutrient anomaly compared to surrounding water, while those descending will be nutrient-depleted. When $\lambda \gg \tau^{-1}$ (fast uptake), the nutrient anomaly is small (Fig.~\ref{fig:3D}B; $\lambda = 1.5 \mathrm{ day }^{-1}$) and in phase with $w$ (Fig.~\ref{fig:3D}A). The vertical nutrient flux $wN'$ at a depth of 115~m (shown in blue-green shades above the cube in Fig.~\ref{fig:3D}B) is small, but positive everywhere, because upward (downward) velocity is correlated with positive (negative) nutrient anomaly.  Conversely, for $\lambda \ll \tau^{-1}$ (slow uptake), the nutrient anomaly is large (Fig.~\ref{fig:3D}C; $\lambda = 0.015 \mathrm{~day}^{-1}$) and out of phase with $w$ (Fig.~\ref{fig:3D}A), resulting in a filamentous distribution of the nutrient as unconsumed nutrient is stirred by the flow. In this case, the vertical flux of nutrient (shown in blue-green shades in Fig.~\ref{fig:3D}C) has large values due to the relatively large nutrient anomalies being advected vertically, but the vertical flux is both positive and negative.  A negative (downward) flux of nutrient occurs when a positive nutrient anomaly is advected downward before it can be taken up by the slowly growing phytoplankton, or a negative nutrient anomaly is transported upward before it is equilibrated to $N_0(z)$. The net flux integrated over the domain is small, as the positive and negative nutrient fluxes cancel each other.  The two cases (Fig.~\ref{fig:3D}B and C) have a similar net nutrient flux -- in one case the local fluxes are small, and in the other, the fluxes are bi-directional and cancel out.  The maximum net nutrient flux is achieved (but not shown in the figure) for a value of $\lambda$ that is intermediate between Fig.~\ref{fig:3D}B and C. Though several numerical modeling studies \cite{mahadevan2000modeling} have reported an enhancement in phytoplankton productivity with an increase in vertical velocity, some \cite{gruber2011eddy} have also reported a reduction in productivity due to the export of unconsumed nutrient as in Fig.~\ref{fig:3D}C.

The spatial distribution of the fluxes depends on the nutrient uptake timescale (Fig. S8). With a slow uptake rate ($\lambda \ll \tau^{-1}$), the spatial distribution of the nutrient anomaly is heavily influenced by stirring in the horizontal and the cascade to small scales driven by lateral stirring \cite{abraham1998generation}. With a fast uptake rate ($\lambda \gg \tau^{-1}$), the nutrient anomaly is present at small spatial scales because the vertical velocity, especially the high frequency component of the vertical velocity, has relatively more small scale variability than the horizontal velocity \cite{mahadevan2002biogeochemical,freilich2020coherent}. 

\section{Lagrangian description of nutrient flux}
In a nonlinear, eddying oceanic flow field, each water parcel has a unique trajectory. As a water parcel moves upward and downward by 1--100 m on timescales of days, it simultaneously moves 1--100 kilometers horizontally. We isolate the effect of vertical motion in our 3D flow field by following water parcels in the Lagrangian frame 
 while tracking their nutrient concentration and nutrient flux by solving $DN/Dt = -\lambda N^\prime$ according to (\ref{eq:original}) on each water parcel trajectory. We experiment with 14 different values of $\lambda$ in the range 0.005 to 50 day$^{-1}$. The time-integrated vertical nutrient flux $\overline{wN'}$ along water parcel trajectories is calculated over 30~days as water parcels move up and down through the nutricline.

\subsection{Theoretical dependence of flux on biological rate}
Theoretically, we can estimate the time-averaged flux $\overline{w N'}$ from  $ \frac{\partial N'}{\partial t} + w \frac{\partial}{\partial z}(N_0(z) + N') = - \lambda N' $, if we assume that the mean free path of water mass trajectories, meaning the average magnitude of depth variation, is small relative to variations in the slope of the background nutrient concentration $\frac{dN_0}{dz}$. This assumption may introduce errors in some situations, but is valid near equilibrium and for small amplitude perturbations (SI Text 2.1). We solve for the nutrient anomaly $N^\prime$ using the integrating factor method (see SI Text 2.2). The vertical flux of nutrient (time averaged over time $t$) is the covariance between the nutrient anomaly and vertical velocity along a given trajectory. It is given by the auto-covariance of $w$ weighted by $\lambda$ as
\begin{equation}
   \overline{w N^\prime} = \frac{dN_0}{dz} \int_{0}^t e^{-\lambda (t-s)}w(s)w(t)ds.
    \label{eq:general_theory}
\end{equation}
Vertical motion transports nutrient due to the gradient $dN_0/dz$ and creates an anomaly in nutrient concentration, which contains a memory of vertical motion for the uptake time ($\lambda^{-1}$) \cite{flierl2002mesoscale}. The vertical velocity auto-covariance matters for a longer period of time for smaller (as compared to larger) values of $\lambda$. 

\subsection{Vertical velocity characteristics and biophysical fluxes}
\begin{figure}[ht!]
\centering
\includegraphics[width=0.95\linewidth]{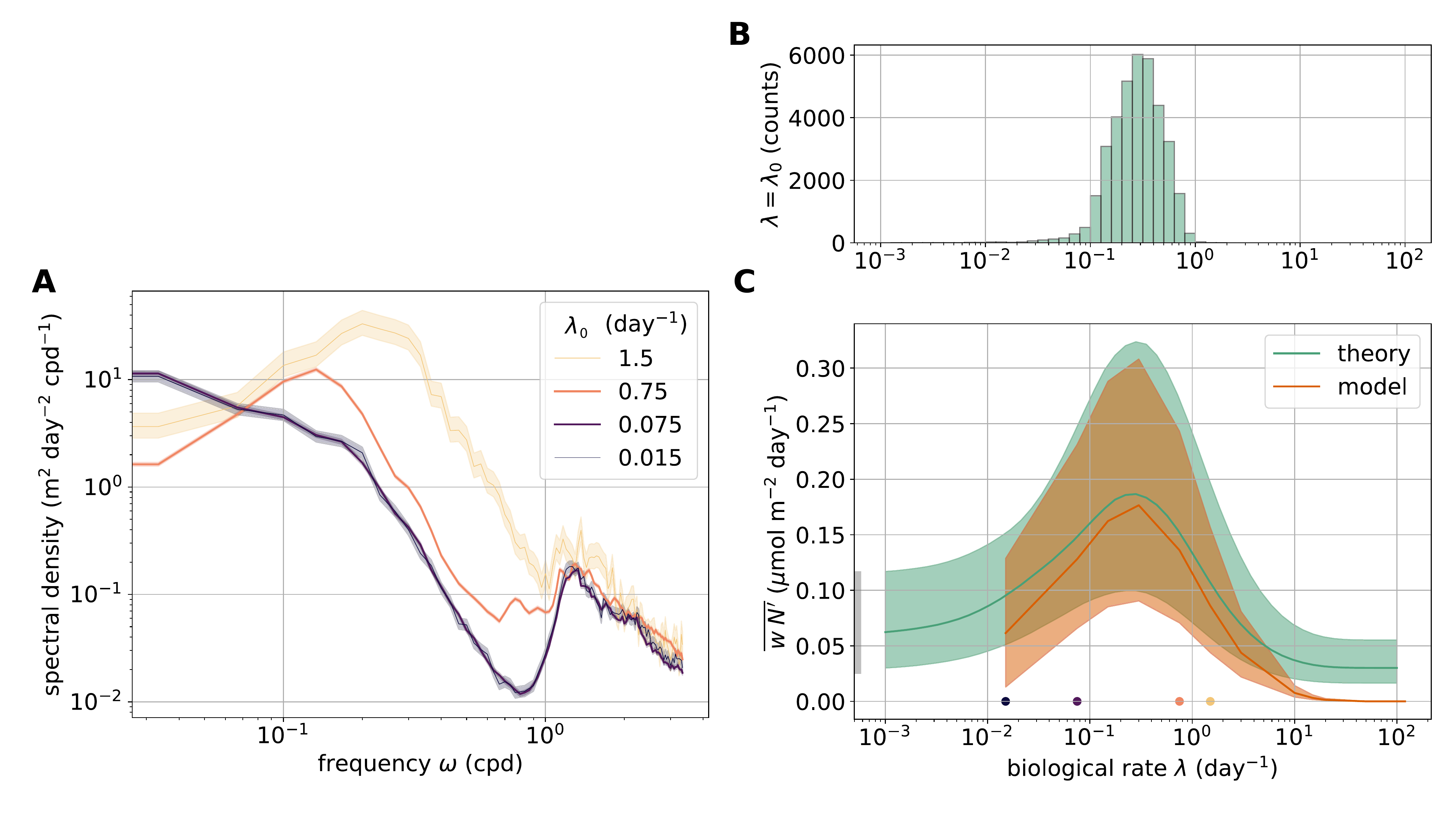}
    \caption{(A) Frequency spectra of the vertical kinetic energy generated from the Lagrangian timeseries. Each curve represents the average of all Lagrangian trajectories whose nutrient nutrient flux is maximized for the  uptake rate $\lambda_0$ indicated in the key. (B) Distribution of rates $\lambda$ that maximizes the nutrient flux. (C) Nutrient flux as a function of uptake rate $\lambda$ on Lagrangian trajectories (orange curve) and theoretical expectation of the nutrient flux from equation (\ref{eq:general_theory}) (green curve). The solid line is the median of all trajectories; shading shows the interquartile range. The grey bar on the ordinate shows the interquartile range of the vertical eddy diffusion in the absence of biological uptake. The dots on the $x$ axis indicate the uptake rates shown in (A).}
    \label{fig:partis}
\end{figure}

For each model trajectory, we identify the value of $\lambda$ that maximizes the flux $\overline{wN'}$. We denote this value as $\lambda_0$. Each water parcel trajectory has a full spectrum of vertical velocity frequencies, but the relative contribution of different vertical velocity frequencies is variable across the domain. The Lagrangian frequency spectrum (Fig.~\ref{fig:partis}A) averaged over all those trajectories on which nutrient flux is maximized for $\lambda=0.075 \mathrm{ day}^{-1}$ (slow uptake) has more power at low frequencies, as compared to the frequency spectrum of trajectories on which nutrient flux is maximized for $\lambda=0.75 \mathrm{ day}^{-1}$ (fast uptake). All trajectories have a peak at the inertial frequency, but the near-inertial oscillations contribute relatively little to the vertical fluxes (Fig. S8).

When we average across all water parcel trajectories in the 3D model, we obtain a maximum in the average Lagrangian nutrient flux as a function of the uptake rate $\lambda$ (Figure \ref{fig:partis}B, orange curve), despite the wide range in the shapes of the vertical kinetic energy spectra of the Lagrangian time series (e.g. Fig.~\ref{fig:partis}A). The uniform uptake rate that maximizes the time-averaged flux in the 3D model is 1/3 day$^{-1}$. The corresponding timescale of 3~days is shorter than the turnover timescale of mesoscale eddies in subtropical gyres, but is characteristic of submesoscale dynamics. 

As a comparison, we calculate the along-trajectory nutrient flux on the full suite of water parcel trajectories according to our theoretical estimate (\ref{eq:general_theory}), which is based on the vertical velocity autocovariance. For any given $\lambda$, the range in the nutrient flux on the different individual trajectories is due to the difference in the vertical velocity frequency characteristics on those trajectories. Along each trajectory, the theoretical expectation of the flux and the flux obtained from the 3D model are consistent (Fig. S2). 

\begin{figure}[ht!]
\centering
\includegraphics[width=0.75\linewidth]{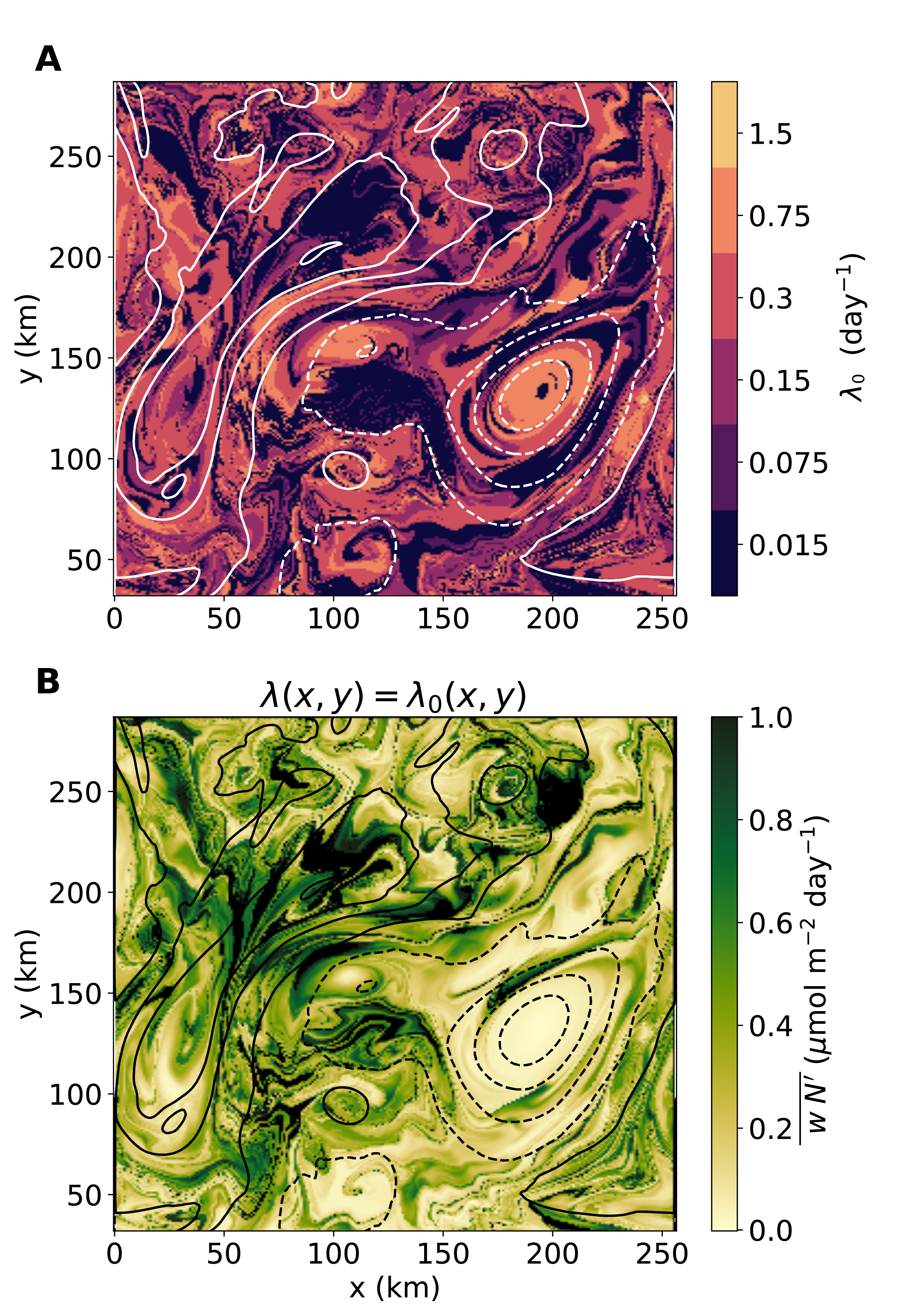}
    \caption{(A) Uptake rate that maximizes the nutrient flux for each trajectory plotted at the trajectory origin location. (B) Nutrient flux at the uptake rate that maximizes the flux. The black points are excluded from the analysis in Figure~\ref{fig:partis} because they leave the nutricline. The contours are density with the dashed contours being lighter densities.}
    \label{fig:comms}
\end{figure}

In the absence of biological uptake, the nutrient flux is given by (\ref{eq:general_theory}) with $\lambda = 0$. This is the Taylor (1922) \cite{taylor1922diffusion} dispersion (Figure \ref{fig:partis}C; grey bar). On average, biological uptake at the rate $\lambda = 1/3 \mathrm{ day}^{-1}$ results in an enhancement of the vertical nutrient flux to three times the flux with no biological uptake. If the community has a uniform uptake rate of 0.3 day$^{-1}$, the median flux is 0.177 $\mu$mol~m$^{-2}$~day$^{-1}$. However, the extreme values of the flux are even larger. If the rate is such that it maximizes the flux locally on each trajectory (Figure \ref{fig:comms}), the median flux is 0.193 $\mu$mol~m$^{-2}$~day$^{-1}$, which is 10\% higher than with uniform $\lambda$.

The difference between a uniform uptake rate and the rate that maximizes the flux on the trajectory relates to the variation in the vertical velocity frequency characteristics (Figure \ref{fig:partis}A). These characteristics relate to the dynamics of the flow field. A spatial distribution of $\lambda_0$ plotted at the origin of each trajectory (Figure \ref{fig:comms}) shows variability that corresponds with the physical features of the flow field. The nutrient flux is maximized by small $\lambda$ (slow uptake) on Lagrangian trajectories that originate in relatively stable regions of the flow, while large values of $\lambda$ (fast uptake) maximize the nutrient flux on trajectories that are in unstable regions or where eddies interact. The process described here is a mechanism for coexistence of distinct biological communities over short spatial scales due to differential growth that is attributable to small scale variability in the vertical velocity \cite{perruche2011effects}. This mechanism differs from lateral stirring of distinct phytoplankton communities, which also results in finescale spatial variability of the community composition \cite{d2010fluid,levy2015dynamical}.

\section{Discussion}
\begin{figure}[ht]
\centering
\includegraphics[width=0.9\linewidth]{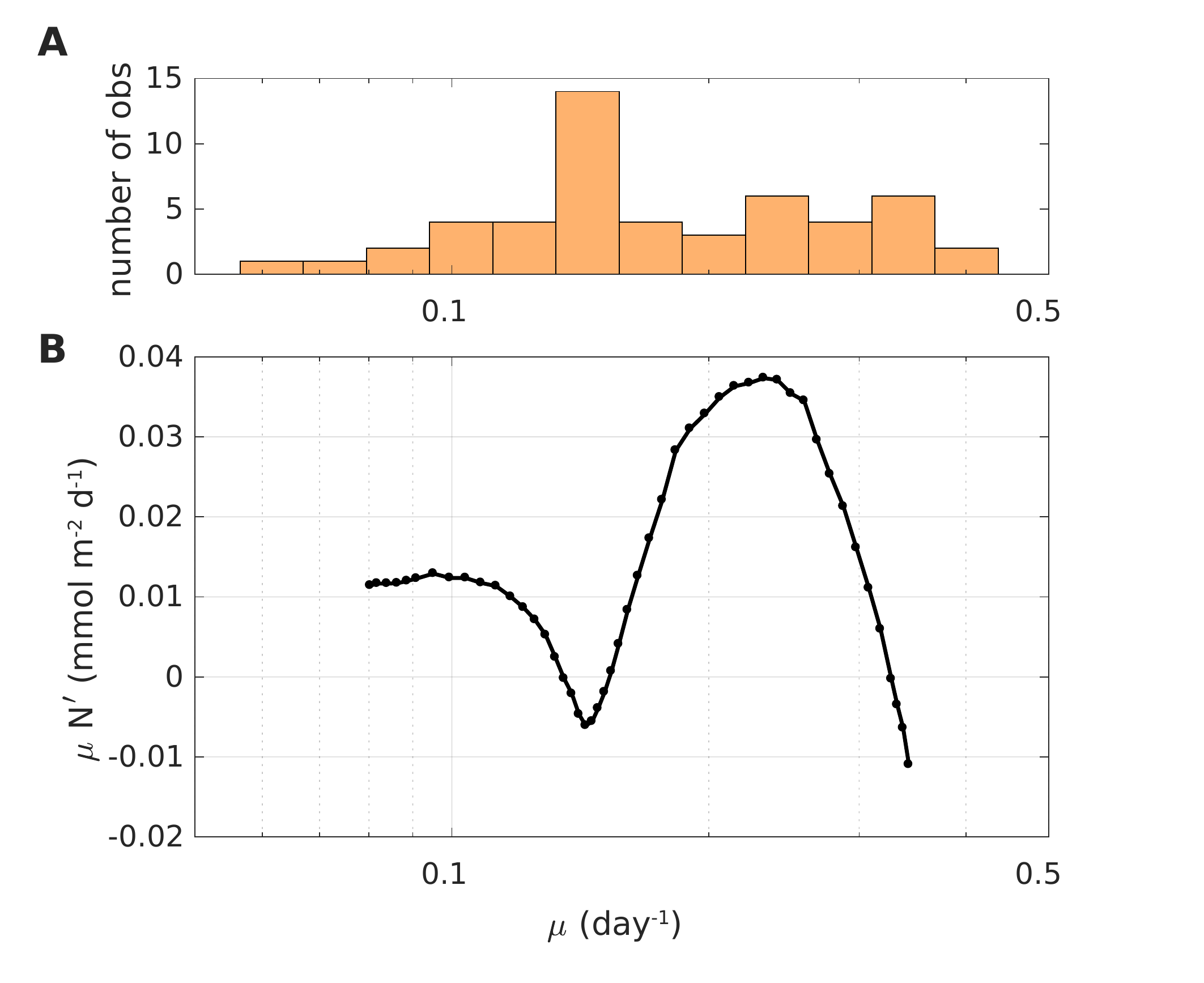}
    \caption{(upper) Histogram of carbon-specific growth rate estimated from primary production and cell counts at the Hawaii Ocean Time-series station ALOHA (lower) Temporal average of biological nutrient flux at 120~m as a function of the carbon specific growth rate at 120~m.}
    \label{fig:obs}
\end{figure}

Observations show that fronts and eddies affect the rate of primary production and the community composition \cite{letelier2000role,rodriguez2001mesoscale,sakamoto2004influence,white2007factors,mcgillicuddy2016mechanisms}. At global scales, phytoplankton community structure has also been observed to depend on local rates of resource supply \cite{maranon2015cell}. At a long-term oceanographic timeseries in the subtropical North Pacific, Station Aloha, there is a six-fold variation in the growth rate in the euphotic zone (Figure \ref{fig:obs}). However, primary production is not linearly correlated with mesoscale eddy variability \cite{barone2019ecological,ferron2021euphotic}. We observe that the average covariance of nutrient anomalies at the base of the euphotic zone and the carbon specific growth rate (Equation \ref{eq:anomaly}) over the 30 year record has a nonlinear dependence on the growth rate. This correspondence with our theoretical prediction suggests that physical processes exert a strong control on the overall nutrient flux and that vertical motions with submesoscale timescales contribute the largest nutrient fluxes. It is notable that a similar biological uptake rate that results in the maximum flux in our 3D model of a subtropical eddy field emerges as the rate with the maximum estimated phytoplankton new production
in the North Pacific subtropical gyre. This rate, $\lambda\approx 0.3 \mathrm{ day}^{-1})$,
is also approximately the average growth rate in the North Atlantic subtropical gyre \cite{goericke1998response}. It is tempting to think that the growth rates of phytoplankton in the ocean are evolved to match the physical rates of supply.

The interrelatedness of the phytoplankton uptake rate, nutrient anomaly, vertical velocity and nutrient supply shown in this model system is applicable to other biogeochemical properties and biological processes such as ocean carbon uptake \cite{mignone2006central,kwon2009impact}, light-limited primary production in the surface mixed layer \cite{dusenberry2000field,taylor2011shutdown}, and the distribution of oxygen, pollutants, and trace elements. This theoretical framework that rigorously defines the relationship between vertical transport characteristics and the biological rate helps to resolve long-standing uncertainties in the role of physical processes in biogeochemical fluxes \cite{gruber2011eddy,liu_ecosystem_2021}.

The model used here neglects the complexity of biological interactions, which are nonlinear and involve multiple timescales. Behaviors such as luxury nutrient uptake among diatoms can also decouple productivity from nutrient uptake. In the case of nonlinear reactions, the theory outlined here will still apply, but analytic solutions are less tractable. In general with a nonlinear biological function $f(N)$ in equation (\ref{eq:original}) the magnitude of the biogeochemical anomaly itself will depend nonlinearly on the biological rate and the timescale may be determined by nonlinear processes such as ecologically-driven oscillations \cite{franks2001phytoplankton,neufeld2012stirring}. Vertical fluxes may also be affected by vertical variations of both biological rates and the vertical velocity magnitude \cite{kahru1983phytoplankton}. In a multi-species system, cross-correlations between components will be a source of spatial and temporal variance \cite{abraham1998generation,falkowski1998biogeochemical,martin2003estimates,morison2020light}. Future work could address the implications of the dependence of flux on biological rate for changes in community composition at the submesoscale while keeping in mind that variations in community composition may also be a result of ecological interactions, but may not alter the net community production \cite{malone1993transient,giovannoni2012seasonality}.


A practical implication of this work for modeling studies that involve physics and biology is the need to give equal importance to resolving and accommodating for similar timescales in both. 
Studies that compare the effect of enhanced submesoscale dynamics at finer model resolutions on primary productivity while holding the biological rates constant may find that the nutrient flux and biological productivity are sensitive to the biological timescales that are selected due to alterations in the mean nutrient gradient \cite{levy2012large,balwada2018submesoscale,resplandy2019effects}. If the vertical velocity processes with timescales on the order of the biological timescales are underrepresented, then there may be a mismatch between the biological and physical processes. 

There is not a clear separation between the mesoscale and submesoscale, but rather a continuum in which the integral timescale of the vertical velocity is affected by eddies. Nonetheless, the
biological uptake timescale that maximizes the nutrient flux 
in our model is within the range of timescales of up- and down-ward motion associated with sub-inertial submesoscale dynamics, corroborating the importance of its contribution for nutrient supply. Mesoscale and submesoscale fluxes are often parameterized as a turbulent diffusion in global models. However, because the timescale of the diffusion feeds back on the nutrient gradient, the biological rate affects the eddy diffusivity coefficient and a diffusive parameterization may not appropriately represent the fluxes, especially when the physical and biological timescales are similar \cite{smith2016effects}. As a consequence, nutrient fluxes and net community production may be underestimated if submesoscale features are not included and the projected changes in ocean biogeochemcial cycles may be highly sensitive to the relationship between the biological and physical models \cite{loptien2019reciprocal,brett2021sensitivity}. The estimation of a Lagrangian flux as suggested here may aid both analysis and parameterization \cite{plumb1979eddy}. Better understanding of the frequency characteristics of submesoscale dynamics \cite{torres2018partitioning,callies2020time} could improve diagnosis of biogeochemical fluxes and ecological community composition. 

\section{Conclusions}
The vertical flux of nutrients is modulated by the coupling between phytoplankton growth rate (nutrient uptake rate) and the frequency (inverse of decorrelation time) of the vertical velocity of water parcels moving up and down through the nutricline. Even linear biological reactions result in a nonlinear dependence of the nutrient flux on the uptake rate $\lambda$, with the maximum flux in a three-dimensional oceanic eddy field  for $\lambda=1/3$~day$^{-1}$ when $\lambda$ is uniform, and an even higher maximum if $\lambda$ varies across water parcels.  At the same time, the vertical flux influences the growth and distribution of phytoplankton. The variability in the spectral characteristics of different water parcel trajectories implies that a range of phytoplankton growth rates is favored to maximize the rate of nutrient supply and new production in an oceanic eddy field. This mechanism of bio-physical coupling supports variability in the structuring of phytoplankton communities on meso- and submeso-scales.

\acknowledgments
We thank Angelicque White for insightful conversations on the interpretation of data obtained from the Hawaii Ocean Time-series HOT-DOGS application; University of Hawai'i at M{\=a}noa. National Science Foundation Award \#1756517.MF and AM were funded by N00014-16-1-3130 (ONR) and MF was also supported by the Martin Fellowship, MIT. 

\bibliography{growth_rate}

\end{document}



\baselineskip24pt


\maketitle

\section{Methods}
\subsection{Physical model}
We simulate an eddying flow field using the Process Study Ocean Model (PSOM) \cite{mahadevan1996nonhydrostatic,mahadevan1996nonhydrostatic1}. The numerical model is initialized with a horizontal buoyancy gradient generated by altering a central temperature and salinity from an Argo profile that was taken in the subtropical gyre on December 11, 2015.  The center of the model domain is at 35$^\circ$N. The model is forced with differential heating and cooling to maintain the initial buoyancy gradient. 
In the vertical, the model has a stretched grid with spacing that ranges from 3~m near surface to 32~m at the bottom. The time step for numerical integration is 432 seconds. The model is analyzed after it has spun up for 120 days, by which time an eddy field is well-developed and the kinetic energy is in equilibrium. The analysis period is 60 days. Lagrangian particles are released across the domain on day 120 in a grid with 1~km spacing at 125~m (81,920 particles are released). The particles are tracked for the 60~day analysis period.

\subsection{Biological model}
We model nutrient concentration ($N$) using equation (1) with 14 distinct values of the uptake rate $\lambda$ in two ways:  on Lagrangian particles and as a tracer that is integrated offline with 3-hourly saved model output. Tracer advection uses a forward Euler scheme and the QUICK advection routines from PSOM. The position of Lagrangian particles is integrated  using an implementation of the Doos and deVries \cite{vries2001calculating} algorithm in Python. We use a hyperbolic tangent function as the background nutrient concentration, which provides a nutricline, or localized gradient in the nutrient profile (SI Fig~\ref{fig:nutrient_obs}), and generates a deep biomass maximum (SI Fig~\ref{fig:chl_obs}). The width of the model nutricline is set by the approximate observed depth range of the deep chlorophyll maximum in the subtropical North Atlantic. While this is a good model of the nitracline, it does not represent the deep nitrate concentration, which increases with depth (and increases with density). However, we assume no production takes place at depth which requires this equilibrium profile to have a slope of zero at depth because we have assumed a constant growth rate  (SI Fig~\ref{fig:nutrient_obs}). 

\subsection{Observations}
We compute the carbon specific growth rate using observations from the Hawaii Ocean Time-series. There are 31 years of monthly observations at station ALOHA, which is located at 22$^\circ$45$^\prime$N, 158$^\circ$00$^\prime$W. Primary production is measured using $^{14}$C incubations for 12 hours at a range of depths within the euphotic zone. The primary production ($\Delta C$) is computed as the difference between light and dark incubations. The carbon specific growth rate is computed \cite{perez_vertical_2006} as
\begin{equation}
    \mu = \ln{\frac{\textrm{C} + \Delta C}{\textrm{C}}}.
\end{equation}
The phytoplankton carbon (C) is estimated from flow cytometric cell counts using the average conversion factors from cells to carbon from \cite{buitenhuis2012picophytoplankton}. The nutrient anomaly is computed as the deviation from the monthly average nutrient concentration at each depth. The production $\mu N^\prime$ is estimated by computing a forward then backward 15 point moving mean of $\mu N^\prime$ evaluated at 120~m sorted by depth integrated $\mu$.

\section{Derivations of flux dependence on nutrient uptake rate}
\subsection{Biological model and discussion of assumptions}
\label{text:assumptions}
The biological model used in the main text can be derived from a nutrient ($N$)- phytoplankton ($P$) model. The evolution of phytoplankton concentration can be described as
\begin{equation}
    \frac{\textrm{d}P(t;z)}{\textrm{d}t} = r(z)N(t;z)P(t;z)-m(z)P(t;z). \label{eq:np-model}
\end{equation}
In this model, the growth rate $r$ and mortality/respiration rate $m$ are both functions of depth. This is a non-linear model that is related to the commonly-used logistic growth equation. We can linearize this model and, assuming a conservative nutrient and phytoplankton system with a depth-dependent total concentration $N_T$ that is partitioned between the nutrient and phytoplankton pools such that $N_T = N+P$  write it in terms of nutrients rather than phytoplankton to obtain the model used in the main text
\begin{equation}
    \frac{\textrm{d}N}{\textrm{d}t} = -\lambda (N-N_0(z)).
\end{equation}
In this model, the parameters $\lambda$ and $N_0(z)$ are both combinations of $r$, $m$, and $N_T$. The uptake rate $\lambda$ is the growth rate in the logistic formulation. We choose to prescribe $\lambda$ as constant and include all depth dependence in $N_0(z)$. This can be done by prescribing a relationship between $r(z)$, $m(z)$, and $N_T(z)$ such that only two of those three variables are assigned independently. The full non-linear form has complex behavior that captures phenomena such as blooms, which are not represented with a linearized growth function.
Using this linearized model, the general equation for the evolution of the nutrient concentration is 
\begin{equation}
    \frac{\partial N}{\partial t}+   \nabla \cdot ({\bf u} N) =   - \lambda ( N- N_0(z) ).
\end{equation}
The major assumption in this equation is that the biological growth function can be appropriately modeled as a linear approximation. This assumption may be violated in certain circumstances. 

\subsection{General theory for deriving the time-averaged nutrient flux on a water parcel}
\label{text:solution}
Since the growth rate in equation (\ref{eq:np-model}) depends only on the vertical position, we can isolate the terms that depend on the vertical coordinate when analyzing the vertical nutrient flux on a water parcel that moves vertically up and down and horizontally in the region of the nutracline. The evolution of nutrient on the water parcel is described by
\begin{equation}
    \frac{\partial N}{\partial t}+w\frac{\partial N}{\partial z} = -\lambda (N-N_0(z))
\end{equation}
Decomposing the nutrient concentration into a background concentration ($N_0$) and an anomaly ($N^\prime$), $N = N_0(z)+N^\prime$, we can derive an equation for the evolution of the nutrient anomaly
\begin{equation}
    \frac{\partial N^\prime}{\partial t}+w\left(\frac{d}{dz} N_0(z)+\frac{\partial N^\prime}{\partial z}\right) = -\lambda N^\prime
\end{equation}
To derive a theoretical estimate of the time-averaged vertical flux of nutrient, $\overline{wN^\prime}$, we make the simplifying assumption that the vertical velocity magnitude is uniform with $z$ and that $N_0(z)$ is a linear function of $z$ within the average fluctuation depth scale of a water parcel. This, combined with the assumption that $\lambda$ is constant with depth, means that we can assume that $N^\prime$ is locally a linear function of depth. Therefore, we can write $\frac{\partial N}{\partial z} = \gamma$ and derive the simplified equation 
\begin{equation}
    \frac{\partial N^\prime}{\partial t}+\gamma w = -\lambda N^\prime.
    \label{eq:nprime}
\end{equation}
We solve for the nutrient anomaly using the integrating factor method. We multiply both sides of the equation 
$\partial_t N^\prime + \lambda N^\prime = -\gamma w$ by $e^{\lambda t}$, recognize that the left hand side is equivalent to $\partial_t (e^{\lambda t}N^\prime )  = -\gamma e^{\lambda t} w $ and integrate both sides to get $e^{\lambda t} N^\prime |_0^t = -\gamma \int_0^t e^{\lambda s} w(s) ds$. Therefore, 
\begin{equation}
    N^\prime(t) = C_0 \int_{0}^t e^{-\lambda (t-s)}w(s) ds+C_1
\end{equation}
The vertical nutrient flux averaged over the time interval $[0, t]$ is then
\begin{equation}
    \overline{w N^\prime}^t = \frac{C_0}{t} \int_{0}^t e^{-\lambda (t-s)}w(s)w(t) ds+C_2.
    \label{eq:theory1}
\end{equation}

\subsection{Idealized physics: Nutrient flux for an oscillating vertical velocity}
\label{text:simplesine}
To build intuition for the dependence of the vertical nutrient flux on the interaction between biological and physical timescales, we examine a case in which the vertical velocity oscillates sinusoidally with angular frequency  $\omega$ and amplitude $a$ as $w=a\sin(\omega t)$.
\subsubsection{Flux dependence on uptake rate}
The time evolution of the nutrient concentration is described by equation (\ref{eq:theory1}) as
\begin{equation}
\frac{\partial N}{\partial t}+a\sin(\omega t)\frac{\partial N}{\partial z} = -\lambda(N-N_0(z)),
\end{equation}
which, in terms of the nutrient anomaly  $N^\prime = N - N_0(z)$, is
\begin{equation}
\frac{\partial N^\prime}{\partial t}+a\sin(\omega t)\bigg(\frac{\textrm{d}N_0}{\textrm{d}z} + \frac{\partial N^\prime}{\partial z}\bigg) = -\lambda N^\prime.
\end{equation}
Making the ansatz $N' = A\sin(\omega t)+B\cos(\omega t)$ gives
\begin{equation}
A\omega \cos(\omega t)-B\omega\sin(\omega t)+a\frac{\textrm{d}N_0}{\textrm{d}z}\sin(\omega t) = -\lambda A \sin(\omega t) - \lambda B \cos(\omega t).
\end{equation}
Collecting the terms, we get
\begin{eqnarray}
A\omega+\lambda B = 0 \\
-B\omega+a\frac{\textrm{d}N_0}{\textrm{d}z}+\lambda A = 0.
\end{eqnarray}
Solving for $A$ and $B$ and substituting into the equation for the nutrient anomaly $N'$ we get
\begin{equation}
N' = \frac{-\lambda a \frac{\textrm{d}N_0}{\textrm{d}z}}{\omega^2+\lambda^2}\sin(\omega t)+\frac{\omega a \frac{\textrm{d}N_0}{\textrm{d}z}}{\omega^2+\lambda^2}\cos(\omega t).
\label{Nprime}
\end{equation}
The flux is then given by integrating the covariance of $w$ and $N^\prime$ over one period
\begin{equation}
\overline{w N^\prime} = \int_0^{\frac{2\pi}{\omega}} w N'dt = \frac{a \frac{\textrm{d}N_0}{\textrm{d}z}}{\omega^2+\lambda^2}\int_0^{\frac{2\pi}{\omega}} a\sin(\omega t)(-\lambda\sin(\omega t)+\omega\cos(\omega t))dt \\
= \frac{-a^2 \frac{\textrm{d}N_0}{\textrm{d}z} \lambda \pi}{\omega^2+\lambda^2}
\label{eq:flux}
\end{equation}
The flux is maximized in this equation for $\lambda = \omega$.

\subsubsection{Correlation between vertical velocity and nutrient anomaly}
There is an important difference between the flux of nutrient and the correlation between the nutrient anomaly and the vertical velocity. The correlation between vertical velocity and nutrient anomaly is normalized by the standard deviation of each of those variables
\begin{equation}
    \rho(wN') = \frac{\left<w N'\right>}{\sqrt{\left<w w\right>}\sqrt{\left< N'N'\right>}}
\end{equation}

\begin{equation*}
    \left< N'N'\right> = \int_0^{2\pi} \bigg(\frac{-\lambda a \gamma}{\omega^2+\lambda^2}\sin(\omega t)\bigg)^2+2\frac{\omega a \gamma}{\omega^2+\lambda^2}\cos(\omega t) \frac{-\lambda a \gamma}{\omega^2+\lambda^2}\sin(\omega t)+\bigg(\frac{\omega a \gamma}{\omega^2+\lambda^2}\cos(\omega t)\bigg)^2 dt
\end{equation*}

\begin{equation}
    \sqrt{\left< N'N'\right>} = \frac{\sqrt{\pi} a \gamma}{\sqrt{\omega^2+\lambda^2}}
    \label{eq:variance}
\end{equation}

\begin{equation}
    \sqrt{\left<w w\right>} = a\sqrt{\pi}
\end{equation}

\begin{equation}
    \rho(w N') = \frac{-a^2 \gamma \lambda \pi}{\omega^2+\lambda^2}\frac{\sqrt{\omega^2+\lambda^2}}{\pi a^2 \gamma} = -\frac{\lambda}{\sqrt{\omega^2+\lambda^2}}
    \label{eq:correlation}
\end{equation}
As expected, the correlation between vertical velocity and nutrient anomaly goes to $-1$ as the growth rate increases, but the flux is small because $N^\prime$ goes to zero.

\subsection{Comparison between 3D model and theory}
\label{text:model_theory}
Across all particle trajectories, there is good agreement between the theoretical expectation of the nutrient uptake rate that maximizes the nutrient flux and the value computed by using a range of uptake rates in equation (\ref{eq:theory}).  The uptake rate that maximizes the flux is computed by solving the biological model on the particle trajectories for a range of different $\lambda$ (Figure \ref{fig:lag_theory_model}). 

The biological model is integrated over a finite, but sufficiently long, time period. One definition of a sufficiently long analysis period is that the
water parcels complete a full cycle of oscillation in the vertical velocity during this period. We can select Lagrangian trajectories where this is true by selecting trajectories where the net displacement over the analysis period is close to zero (less than 2.5~m over 30~days). The excluded trajectories have a maximize their nutrient flux at longer timescales. When we include all trajectories, the flux at slow uptake rates increases as does the flux in the absence of biological uptake (Figure \ref{fig:rate_all}B). There is also a long tail on the distribution of rates that maximize the flux (Figure \ref{fig:rate_all}A).

One of the major assumptions about the theoretical model is that there is relatively little variation in the vertical velocity and background nutrient concentration gradient over the mean free path of the water parcel trajectory, which allows us to assume that the nutrient anomaly is linear with depth. Solutions could be obtained for other nutrient anomaly profiles as well, but are not derived here. Examining histograms of the nutrient anomaly as a function of depth, we see that this assumption is generally met (Figure \ref{fig:nprime_zprime}). With the slower growth rates, the nutrient anomaly has the same depth dependence as the background profile $N_0(z)$. There are just a few extreme cases with faster growth rates where this assumption does not hold. 

\subsection{Background nutrient profile in model}
The background nutrient profile is based on observations. Due to the linear nature of the model and the modeling choice that $\lambda$ is constant with depth, all production will occur where there is a gradient in the background nutrient profile. We ensure that the production is localized in the region of the deep chlorophyll maximum at BATS by setting up the function $N_0(z)$ such that its derivative has the same shape as the chlorophyll profile at BATS (Figure \ref{fig:chl_obs}). 

The resulting nutrient profile $N_0(z)$ is similar to the profile in the subtropical gyres, although it is notable that the nutricline at the Hawaii Ocean Timeseries (HOT) Station Aloha is steeper than the nutricline at BATS (Figure \ref{fig:nutrient_obs}). 

If we reduce the slope of the background nutrient profile, the variations in the magnitude of the vertical velocity with depth become more important and the nutrient anomaly is less linear with depth. For example, we can reduce the magnitude of the background nutrient anomaly by changing the value of $A$ in the function $N_0(z) = A\left(1-\tanh\left(\frac{z-N_\textrm{cline}}{30}\right)\right)$. In most model simulations, we use $A = 2.5$ for a  nutrient concentration of 5~$\mu$-mol m$^{-3}$ at a depth of 200~m. To reduce the slope of the nutricline, we use $A = 1$ which only gives a  nutrient concentration of 2~$\mu$-mol m$^{-3}$ at a depth of 200~m.. With a smaller value of $A$, the average flux across all water parcels displays a similar shape as in Figure~\ref{fig:rate_all}, but the fluxes are smaller in magnitude, with negative values of flux at high uptake rates and notably greater variation amongst trajectories (Figure~\ref{fig:smallslope}). 

\subsection{Relationship between autocovariance and optimal flux}
    \label{eq:theory}
The autocovariance relates the vertical velocity and flux characteristics. The effects of longer timescales (larger $\tau$) are amplified in determining the uptake rate that maximizes the flux. While there are many possible autocovariance functions that result in an optimal uptake rate that maximizes the flux, in general the first zero crossing of the autocovariance is at a shorter time lag with a faster uptake rate (Figures \ref{fig:acov} and \ref{fig:first_zero}). The details of the autocovariance function are important for determining the exact rate $\lambda$ where the flux is maximum. The autocovariance function is in the time domain and is related to the vertical velocity spectrum in the frequency domain (Fig. 4) according to the Wiener–Khinchine theorem.

\section{Patchiness as a function of uptake  rate}
\label{chap:appendix_spectrum}

The relationship between the spatial and temporal characteristics of the vertical kinetic energy (VKE) can be examined in the wavenumber-frequency space (Figure \ref{fig:cospectra}A); the VKE
lies on the non-dispersive line $\omega = ck$ with phase speed 17 km/day. This relatively fast phase speed is likely due to the influence of strong mesoscale eddies on the vertical velocity distributions. For large $\lambda$ (fast nutrient uptake), the spatial and temporal scale of nutrient uptake ($\lambda N'$) is similar to the vertical velocity (Figure \ref{fig:cospectra}B). However, for small $\lambda$ (slow uptake) the nutrient uptake occupies a different region of the wavenumber-frequency space than the vertical kinetic energy. 

Our results highlight the existence of two types of patchiness that correspond to the extremes of the distribution of flux as a function of biological uptake rate. Nutrient anomalies are injected at the scale of the vertical velocity variability, which for large $\lambda$ results in small-scale patchiness in the nutrient flux and new production. This is the type of patchiness discussed in \cite{mahadevan2002biogeochemical}. By contrast, for small $\lambda$ (slow uptake), the nutrient fluxes are preferentially at lower frequencies and larger spatial scales and the tracer anomalies are stirred by the horizontal velocity field to smaller scales resulting in a filamentous distribution of nutrient and biomass \cite{abraham1998generation}. However, the filaments in nutrient anomaly do not propagate in the same way as the vertical velocity, so the vertical velocity and tracer anomaly are no longer in phase and the tracer flux is weak (Figure \ref{fig:cospectra}).

\section{Observational support of flux dependence on phytoplankton growth}
The North Pacific subtropical gyre is stratified and nutrient limited throughout the year. In this region, productivity has been observed to be influenced by mesoscale eddies \cite{letelier2000role,sakamoto2004influence,white2007factors}. Time series observations from the Hawaii Ocean Time Series include over 30 years of monthly observations of nitrate concentration and $^{14}$C uptake in daytime incubation \cite{karl1996hawaii}. The $^{14}$C incubation measures a physiological quantity of the plankton community that is between the gross and net carbon uptake \cite{marra2002approaches}. Although this quantity is not exactly the uptake rate and is not correlated with the division rate or net community production, it may be a reasonable approximation of the uptake of carbon by photosynthesis \cite{milligan2015advancing,ferron2021euphotic}. However, as an incubation method is a prone to bottle effects \cite{quay2010measuring}. In the observational comparison in Figure~6, we average over all data sorted by uptake rate using a forward-backward 15 point running mean because the monthly station resolution is not sufficient to capture submesoscale frequencies. By averaging over all available data, we are able to include 15 data points per uptake (growth) rate. However, by doing the averages this way we risk the inclusion of correlations with growth rate that overwhelm the signal of the process that we have proposed. One of the major potential issues could be the seasonal cycle. There is some seasonality to the system with the highest temperatures and the highest growth rates observed during the summertime \cite{church2006temporal}, although the seasonality in carbon-specific growth rate at 120~meters is weak (Figure \ref{fig:monthly}). The mesoscale eddy kinetic energy measured from satellites has a maximum in April-May \cite{chen2010mesoscale}. The higher summertime eddy activity could increase the observed nutrient anomaly during times when the growth rate is also high, but we find instead that there are relatively small nutrient anomalies during the summer. The diffusive transport of carbon is constant throughout the year although the lateral transport may vary throughout the year \cite{keeling2004seasonal}. However, the mixed layers are deeper in the wintertime, potentially allowing submesoscale features to reach into the nutricline during the winter. We do not present any error bars because the theory presented is for the average flux. We show all data in Figure \ref{fig:scatter_obs}. There are 2 extreme values of nutrient flux that are due to large nutrient anomalies. These are removed before computing the average curve. 
\bibliography{growth_rate}
\bibliographystyle{ScienceAdvances}

\begin{figure}[htb]
\centering
\includegraphics[width=0.75\linewidth]{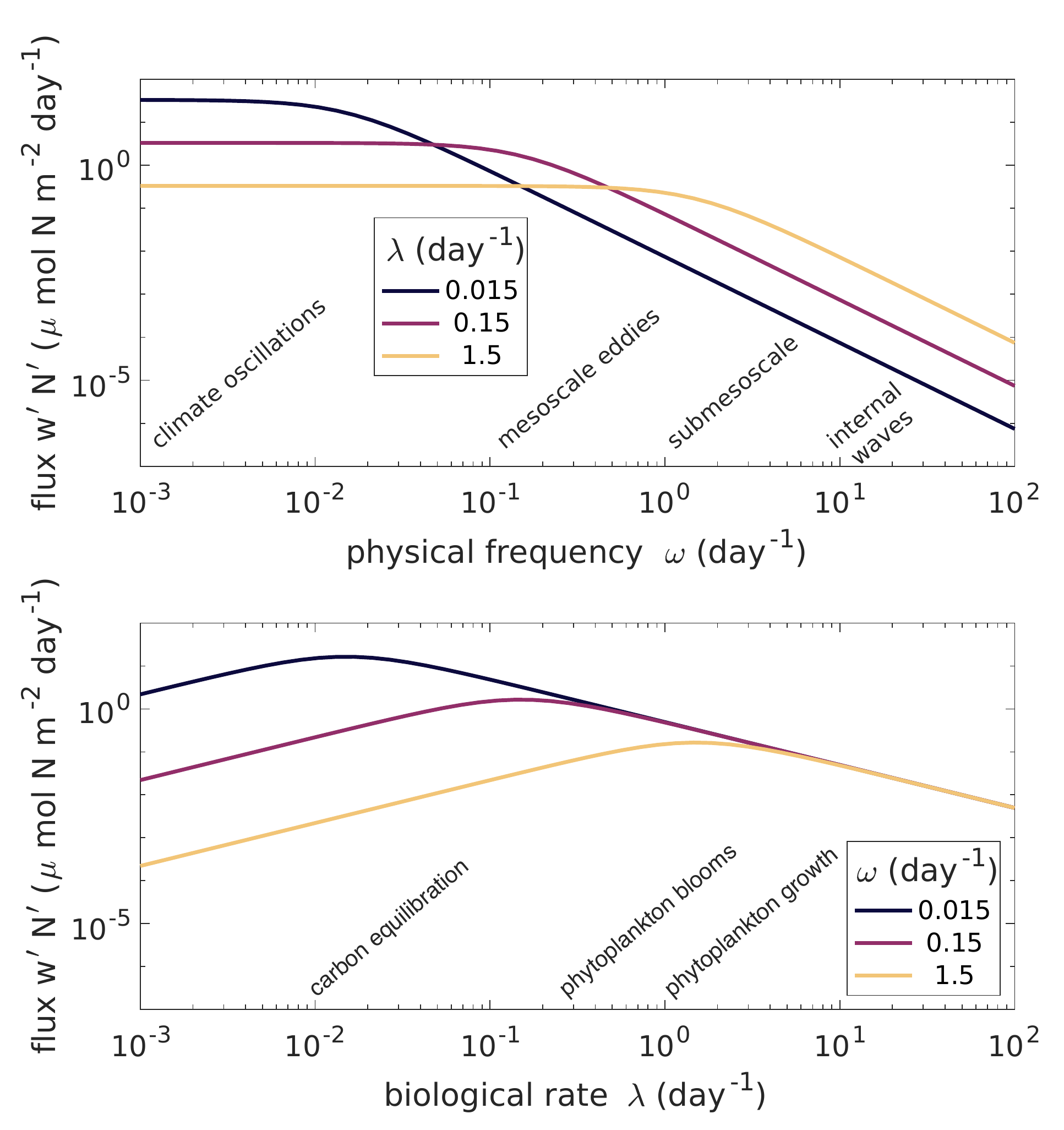}
    \caption{(top) Vertical nutrient flux as a function of growth rate for a range of vertical velocity frequencies. (bottom) Vertical nutrient flux as a function of vertical velocity frequency ($\omega = 2\pi\tau^{-1}$) for a range of growth rates. The vertical flux is computed using the idealized model in equation \ref{eq:flux}.}
    \label{fig:schematic1}
\end{figure}

\newpage
\begin{figure}
    \centering
    \includegraphics[width = 0.75\textwidth]{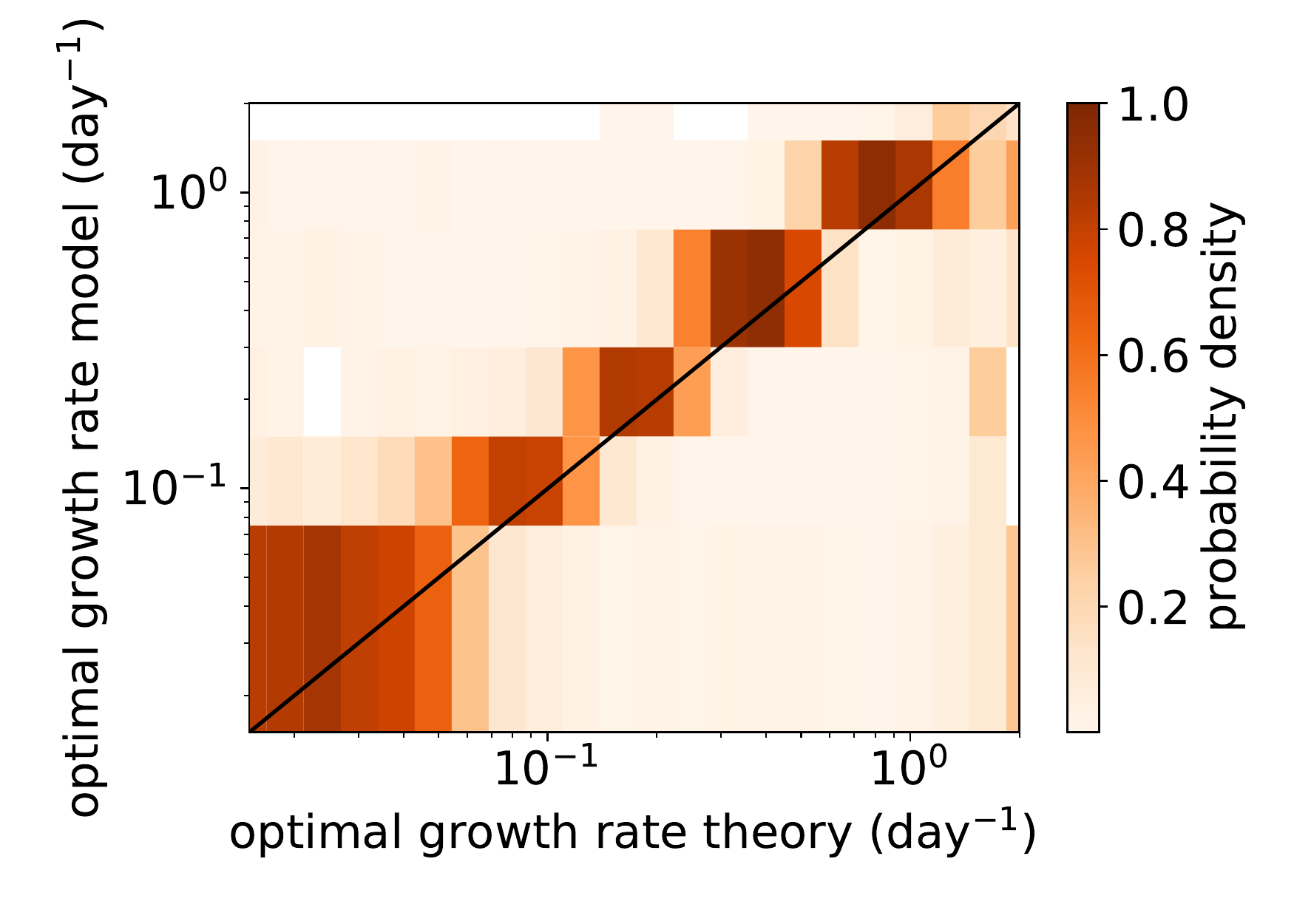}
    \caption{Histogram of the optimal nutrient flux calculated using equation \ref{eq:theory} (``theory'') and on particles in the 3D model. The frequencies are normalized by the number of particles with a given optimum in the theoretical calculation.}
    \label{fig:lag_theory_model}
\end{figure}

\newpage
\begin{figure}
    \centering
    \includegraphics[width = 0.75\textwidth]{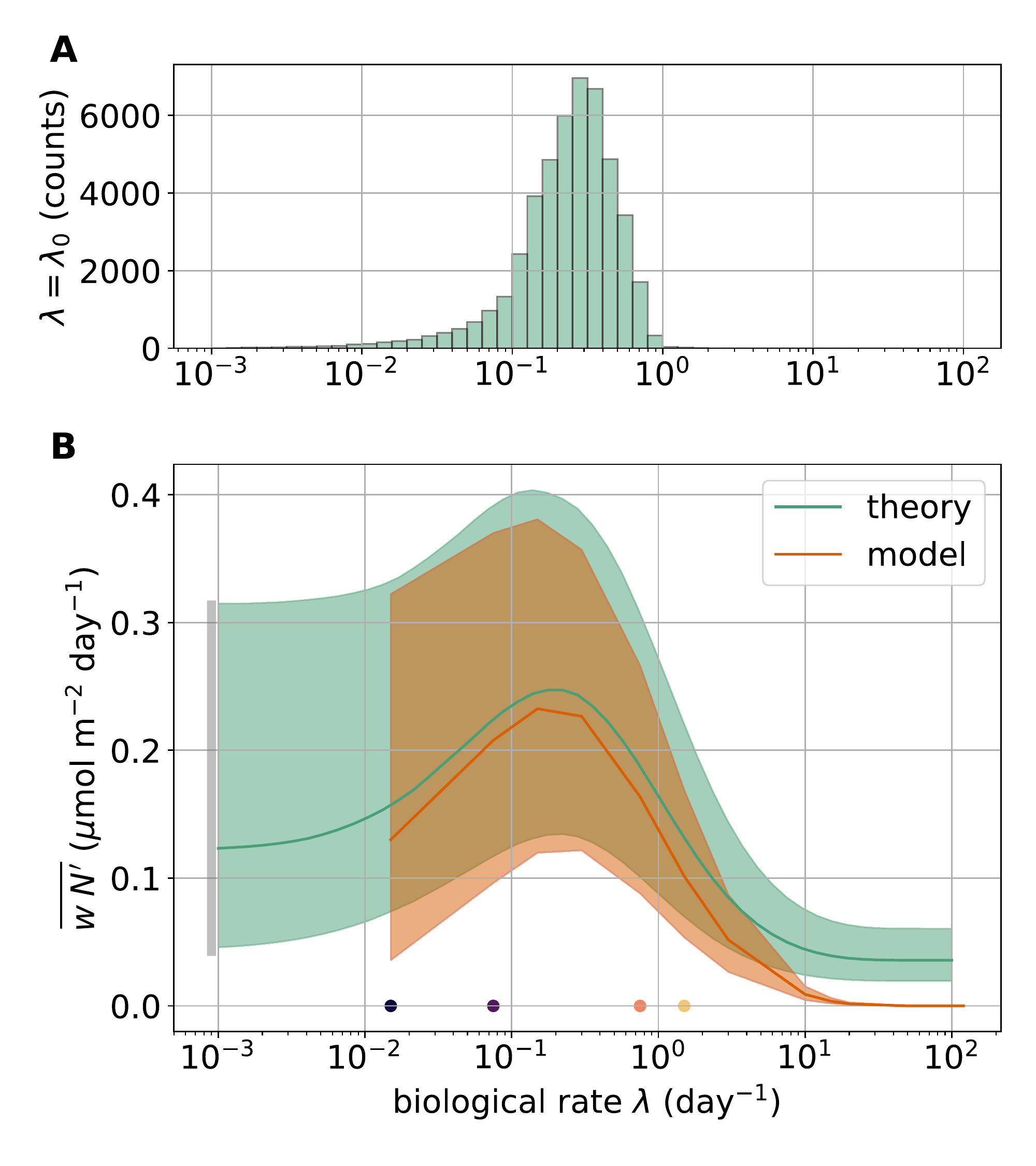}
    \caption{(A) Distribution of rates $\lambda$ that maximizes the nutrient flux. (B) Nutrient flux as a function of uptake rate on Lagrangian trajectories (orange curve) and theoretical expectation of the nutrient flux from equation 3 (green curve). The solid line is the median of all trajectories shading shows the interquartile range. The grey bar on the ordinate shows the interquartile range of the vertical eddy diffusion in the absence of biological uptake.}
    \label{fig:rate_all}
\end{figure}

\newpage
\begin{figure}
    \centering
    \includegraphics[width=\textwidth]{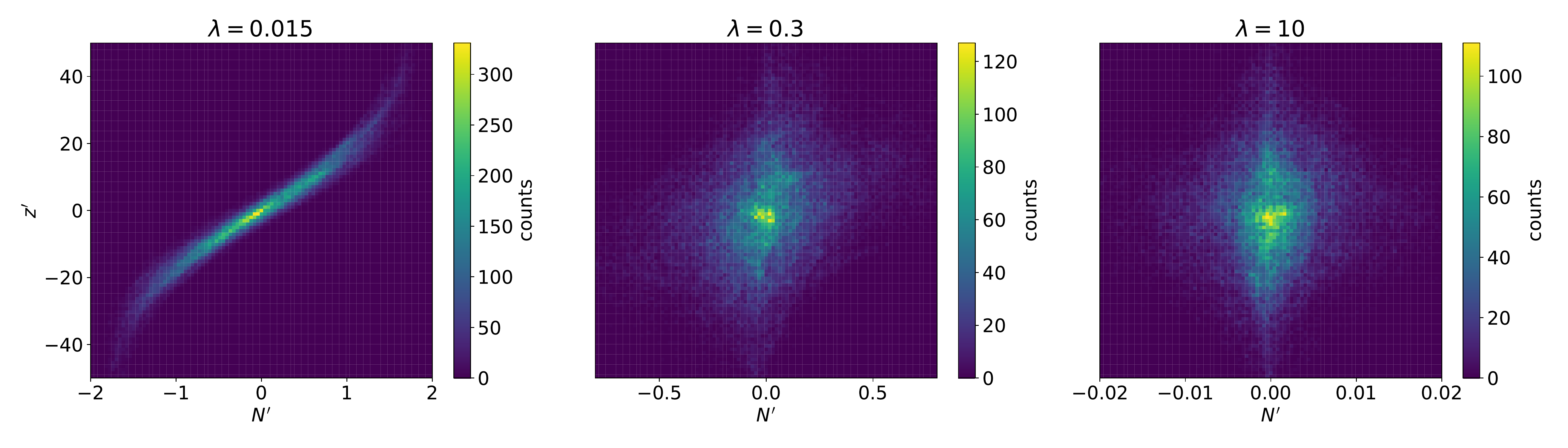}
    \caption{Nutrient anomaly as a function of the depth anomaly (vertical distance from z = -155~m) for a range of biological uptake rates.}
    \label{fig:nprime_zprime}
\end{figure}

\newpage
\begin{figure}
    \centering
    \includegraphics[width = 0.75\textwidth]{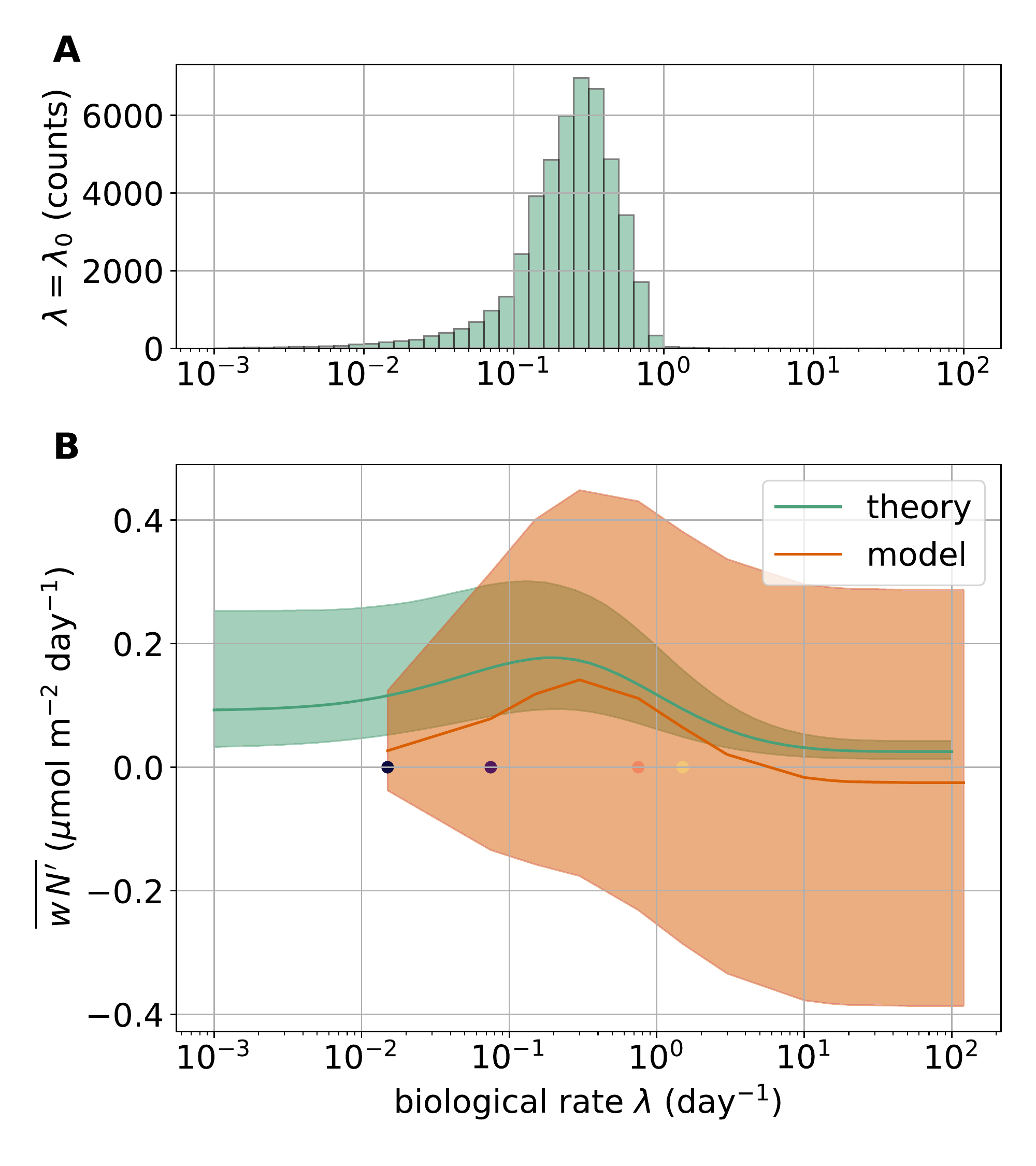}
    \caption{(A) Distribution of rates $\lambda$ that maximizes the nutrient flux. (B) Nutrient flux as a function of biological uptake rate on Lagrangian trajectories (orange curve) and theoretical expectation of the nutrient flux from equation 3 (green curve). The solid line is the median of all trajectories shading shows the interquartile range. The grey bar on the ordinate shows the interquartile range of the vertical eddy diffusion in the absence of biological uptake.}
    \label{fig:smallslope}
\end{figure}

\newpage
\begin{figure}
    \centering
    \includegraphics[width = 0.75\textwidth]{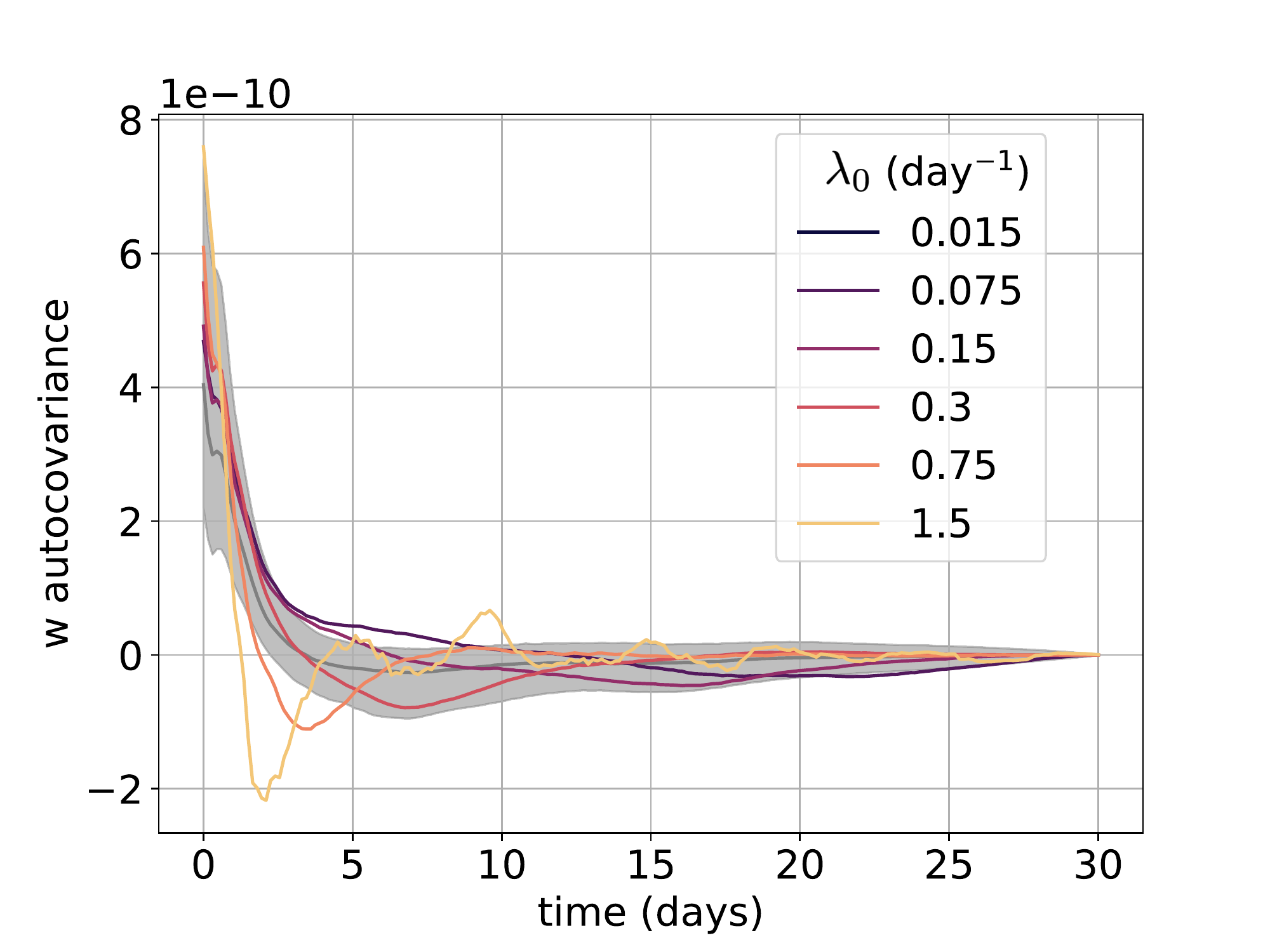}
    \caption{Vertical velocity autocorrelation as a function of time. The grey curve shows the median and interquartile range of all trajectories. The colored lines show the median autocorrelation across all trajectories with a given biological uptake rate the maximizes the flux.}
    \label{fig:acov}
\end{figure}

\newpage
\begin{figure}
    \centering
    \includegraphics[width = 0.75\textwidth]{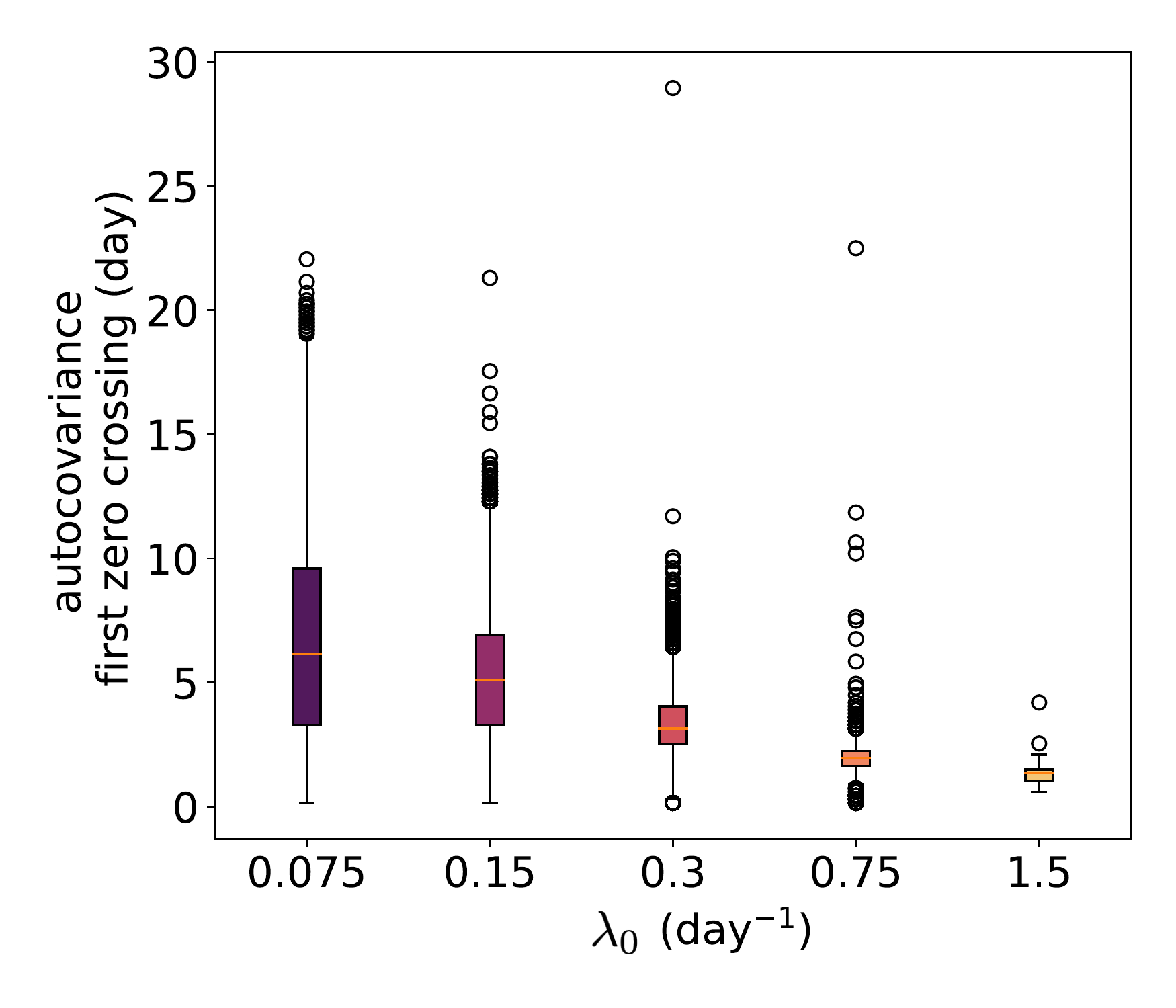}
    \caption{First zero crossing of the vertical velocity autocorrelation. The biological uptake rate on the abscissa is the rate the maximizes the vertical flux. The box plots show the distributions of first zero crossings across all trajectories.}
    \label{fig:first_zero}
\end{figure}

\newpage
\begin{figure}[ht]
\centering
\includegraphics[width=0.9\linewidth]{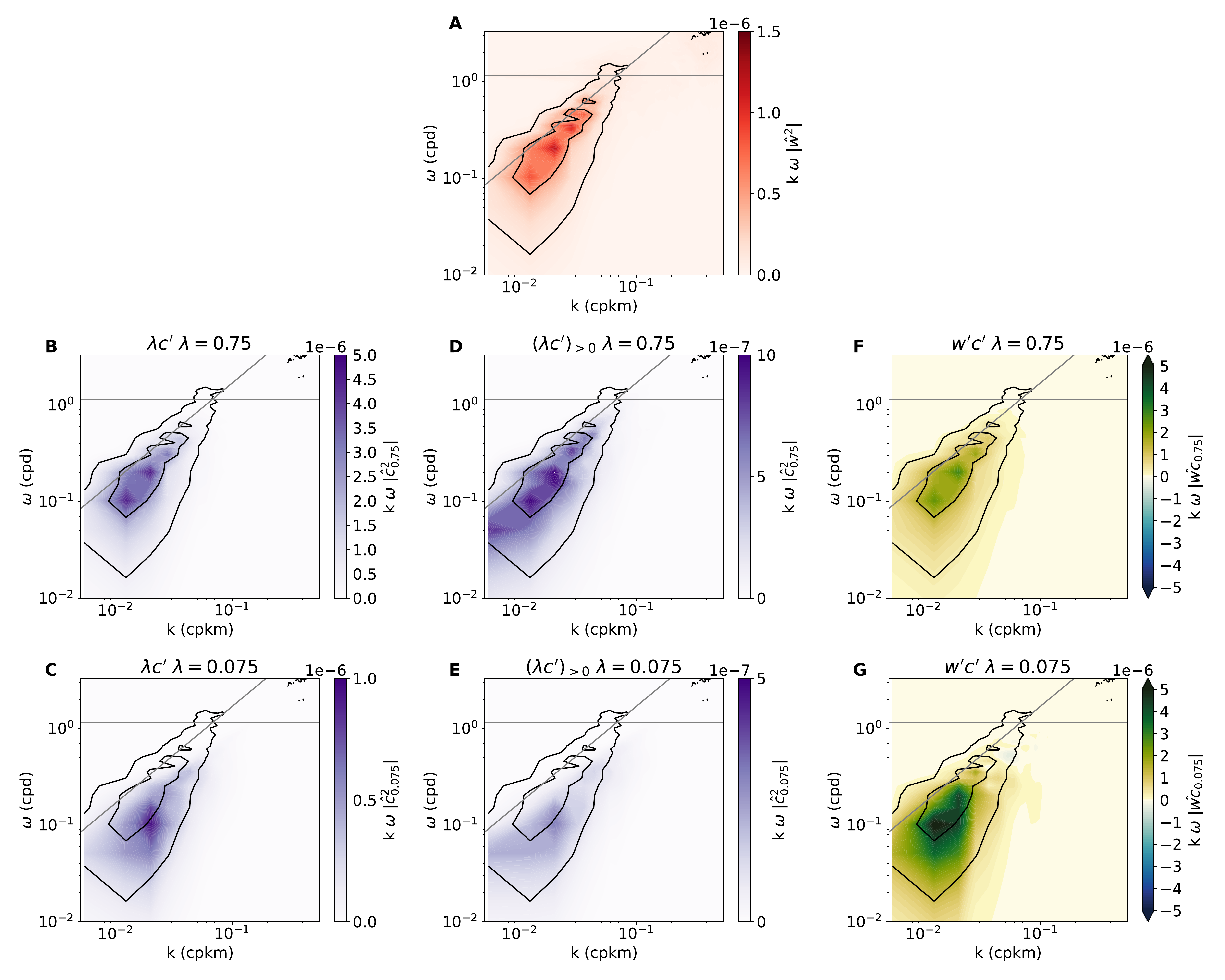}
    \caption{(A) Wavenumber-frequency spectrum of the vertical kinetic energy at 117~meters. (B, C) Wavenumber-frequency of the biological reaction term ($\lambda c^\prime$) at 117~meters with (B) a relatively fast growth rate ($\lambda = 0.75$) and (C) a relatively slow growth rate ($\lambda = 0.075$). (D, E) Wavenumber-frequency of the new production, the positive part of $\lambda c^\prime$, at 117~meters with (D) a relatively fast growth rate ($\lambda = 0.75$) and (E) a relatively slow growth rate ($\lambda = 0.075$). (F, G) Wavenumber-frequency of the nutrient flux ($\lambda c^\prime$) at 117~meters with (F) a relatively fast growth rate ($\lambda = 0.75$) and (G) a relatively slow growth rate ($\lambda = 0.075$). On each panel the grey horizontal line is at the inertial frequency. The grey diagonal line is a non-dispersive line with a phase speed of 17~km/day. The black contours are level curves of the vertical velocity wavenumber-frequency spectrum.}
    \label{fig:cospectra}
\end{figure}

\newpage
\begin{figure}
    \centering
    \includegraphics[width = 0.75\textwidth]{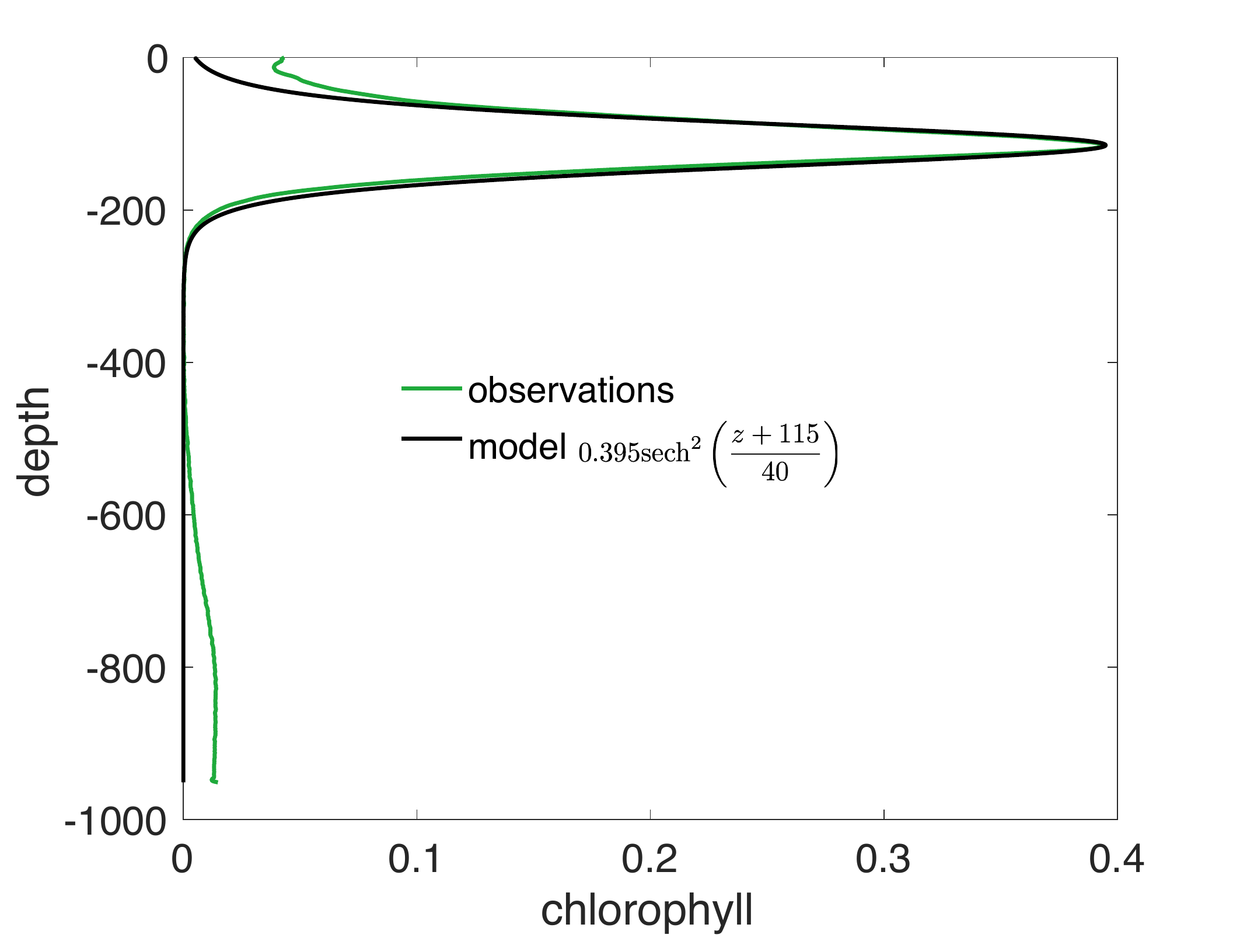}
    \caption{Chlorophyll as a function of depth observed at BATS in June (green) and $\frac{\textrm{d}N_0}{\textrm{d}z}$}
    \label{fig:chl_obs}
\end{figure}

\newpage
\begin{figure}
    \centering
    \includegraphics[width = 0.75\textwidth]{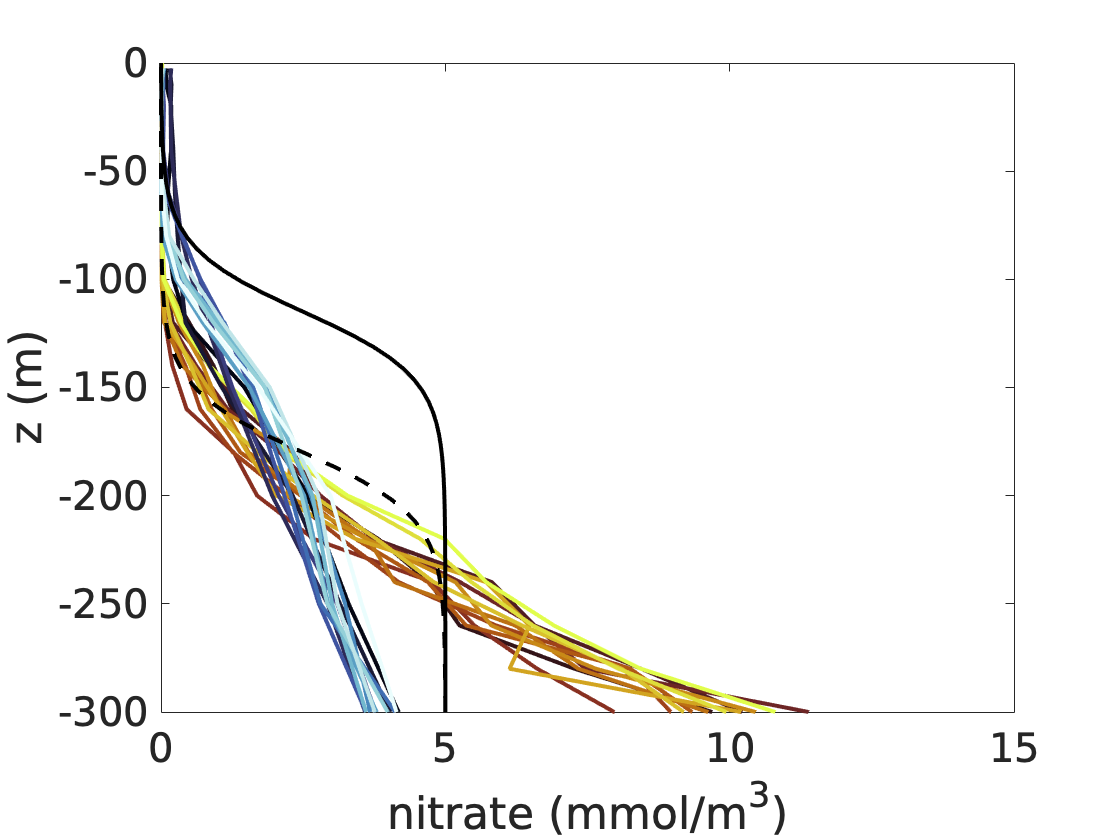}
    \caption{Model and observational nutrient profiles. The model equilibrium profile is the black line and the dashed line is a shifted profile to match the nutricline depth at HOT. The observational profiles are monthly mean profiles from the Bermuda Atlantic Time Series (blues) and Hawaii Ocean Timeseries (oranges). The colors vary from dark to light with month from January to December. The model profile has no gradient below the maximum euphotic depth so that there is no productivity below this depth, due to the model simplifications. The model has a larger nutrient gradient in the nutricline than do the observations at BATS.}
    \label{fig:nutrient_obs}
\end{figure}

\begin{figure}
    \centering
    \includegraphics[width = \textwidth]{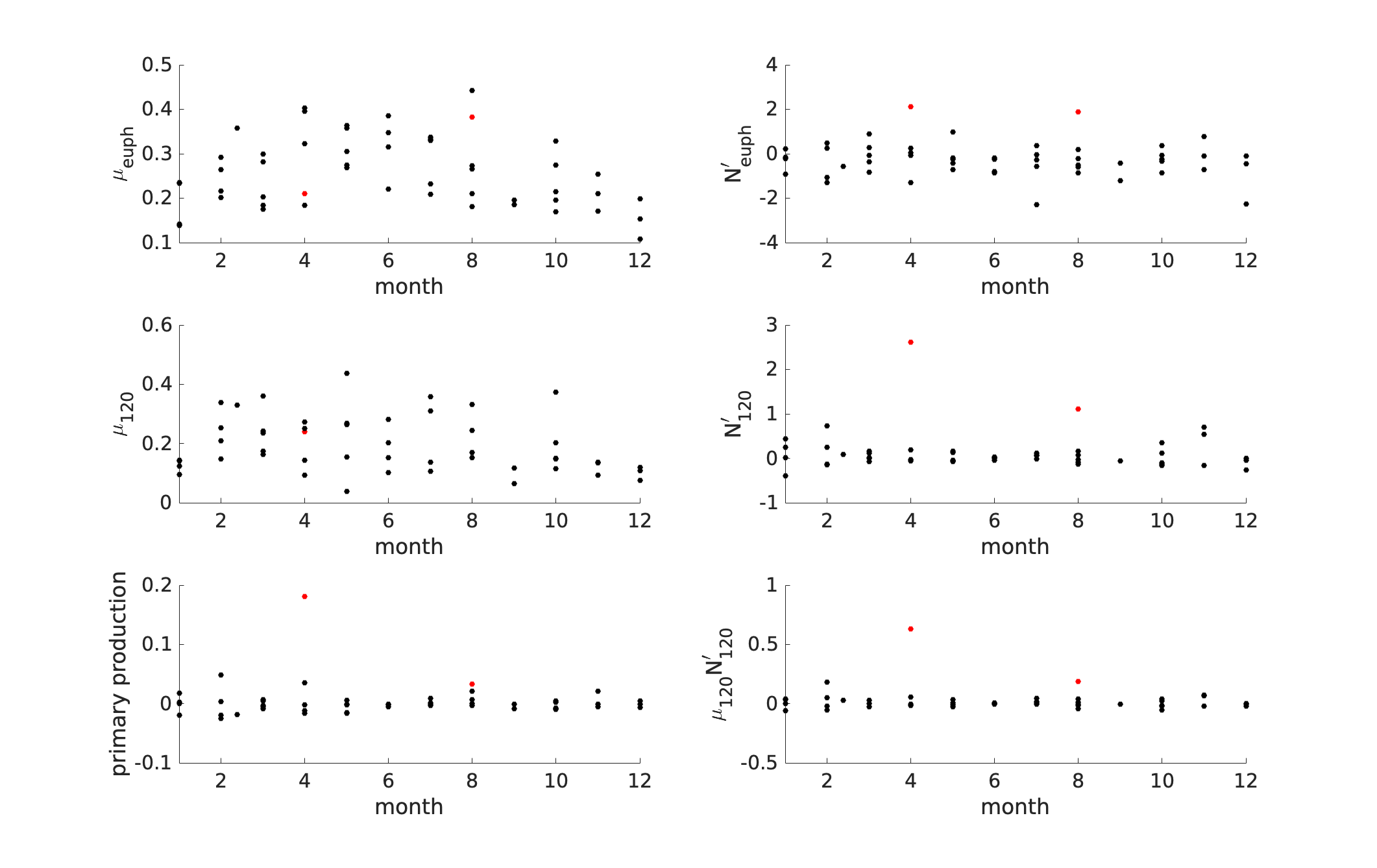}
    \caption{From top to bottom, left to right, measured and computed quantities as a function of month, euphotic zone average carbon specific growth rate, carbon specific growth rate at 120~meters, euphotic zone average primary production, average nutrient anomaly in the euphotic zone, average nutrient anomaly at 120~m, nutrient flux calculated as $\mu N^\prime$ at 120~m. The black dots show the most extreme values of nutrient anomaly. }
    \label{fig:monthly}
\end{figure}

\begin{figure}
    \centering
    \includegraphics[width = \textwidth]{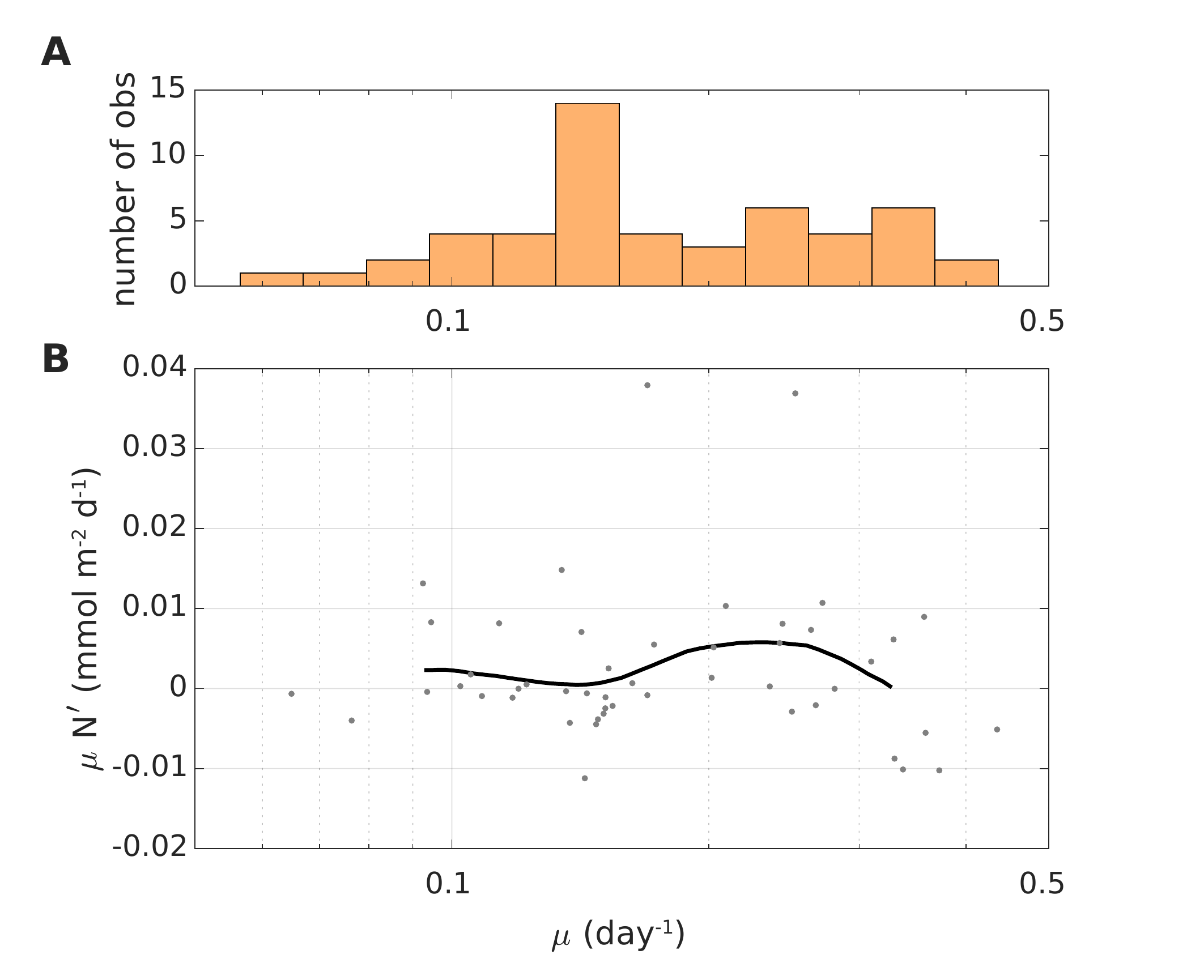}
    \caption{Scatter plot of all observations of the estimated nutrient flux ($\mu N^\prime$) as a function of carbon-specific growth rate ($\mu$). }
    \label{fig:scatter_obs}
\end{figure}
